\documentclass[useAMS,usenatbib,usegraphicx]{mn2e}
\usepackage{times}
\usepackage[total={17.8cm,24.0cm},centering]{geometry} 

\def\aj{AJ}             
\def\araa{ARA\&A}       
\def\apj{ApJ}           
\def\apjl{ApJ}          
\def\apjs{ApJS}         
\def\apss{Ap\&SS}       
\def\aap{A\&A}          
\def\aaps{A\&AS}        
\def\mnras{MNRAS}       
\def\pasp{PASP}         
\def\pasj{PASJ}         

\def\oiii{[O\,{\sc iii}]}
\def\halpha{\hbox{H$\alpha$}}
\def\hbeta{\hbox{H$\beta$}}
\def\ni{\hbox{[N\,{\sc i}]}}
\def\nii{\hbox{[N\,{\sc ii}]}}

\def\hii{\hbox{H\,{\sc ii}}}

\def\Msun{${\rm M_{\odot}}$}
\def\kms{$\mbox{km s}^{-1}$}
\newcommand{\sauron}{{\texttt {SAURON}}}
\newcommand{\Xsauron}{{\texttt {XSAURON}}}
\newcommand{\gmos}{{\texttt {GMOS}}}
\newcounter{subfigure}

\title[The \sauron\ project~-~VII]
{The SAURON project - VII. Integral-field absorption and emission-line 
kinematics of 24 spiral galaxy bulges}

\author[Falc\'on-Barroso et al.]
{Jes\'us Falc\'on-Barroso$^{1}$\thanks{jfalcon@strw.leidenuniv.nl}, 
Roland Bacon$^{2}$, Martin Bureau$^{3}$, Michele Cappellari$^{1}$, 
\newauthor 
Roger L. Davies$^{3}$, P.T. de Zeeuw$^{1}$, Eric Emsellem$^{2}$, 
Kambiz Fathi$^{4}$, Davor Krajnovi\'c$^{3}$,
\newauthor
Harald Kuntschner$^{5}$, Richard M. McDermid$^{1}$, Reynier F. Peletier$^{6,7}$,
Marc Sarzi$^{3}$\\
$^1$Sterrewacht Leiden, Niels Bohrweg~2, 2333~CA Leiden, The Netherlands\\
$^2$Centre de Recherche Astrophysique de Lyon - Observatoire, 9~Avenue Charles
Andr\'e, 69561 Saint Genis Laval, France\\
$^3$Sub-Department of Astrophysics, University of Oxford, Denys Wilkinson
Building, Keble Road, Oxford OX1 3RH, United Kingdom\\
$^4$Rochester Institute of Technology, 84 Lomb Memorial Dr., Rochester, NY
14623-5603, USA\\
$^5$Space Telescope European Coordinating Facility, European Southern
Observatory, Karl-Schwarzschild-Str~2, 85748 Garching, Germany\\
$^6$Kapteyn Astronomical Institute, P.O. Box 800, 9700 AV Groningen,
The Netherlands\\
$^7$School of Physics \& Astronomy, University of Nottingham, Nottingham, NG7
2RD, United Kingdom}

\pagerange{\pageref{firstpage}--\pageref{lastpage}} \pubyear{2005}

\begin{document}
\maketitle
\label{firstpage}

\begin{abstract}
We present observations of the stellar and gas kinematics for a representative
sample of 24 Sa galaxies obtained with our custom-built integral-field
spectrograph \sauron\ operating on the William Herschel Telescope. The data have
been homogeneously reduced and analysed by means of a dedicated pipeline. All
resulting datacubes were spatially binned to a minimum mean signal-to-noise 
ratio of 60 per spatial and spectral resolution element. Our maps typically 
cover the bulge dominated region. We find a significant fraction of kinematically
decoupled components (12/24), many of them displaying central velocity
dispersion minima. They are mostly aligned and co-rotating with the main body of
the galaxies, and are usually associated with dust discs and rings detected
in unsharp-masked images. Almost all the galaxies in the sample (22/24) contain
significant amounts of ionised gas which, in general, is accompanied by the
presence of dust. The kinematics of the ionised gas is consistent with circular
rotation in a disc co-rotating with respect to the stars. The distribution of 
mean misalignments between the stellar and gaseous angular momenta in the sample 
suggest that the gas has an internal origin. The \oiii/\hbeta\ ratio is usually 
very low, indicative of current star formation, and shows various morphologies 
(ring-like structures, alignments with dust lanes or amorphous shapes). The star 
formation rates in the sample are comparable with that of normal disc galaxies. 
Low gas velocity dispersion values appear to be linked to regions of intense 
star formation activity. We interpret this result as stars being formed from 
dynamically cold gas in those regions. In the case of NGC\,5953, the data 
suggest that we are witnessing the formation of a kinematically decoupled 
component from cold gas being acquired during the ongoing interaction with 
NGC\,5954.
\end{abstract}

\begin{keywords}
galaxies: bulges -- galaxies: spiral -- galaxies: kinematics and 
dynamics -- galaxies: evolution -- galaxies: formation -- galaxies: ISM 
\end{keywords}

\section{Introduction}
\label{sec:intro}
We are carrying out a study of the two-dimensional kinematic and stellar
population properties of a representative sample of 72 early-type galaxies,
primarily using integral-field observations obtained with \sauron\
\citep[hereafter Paper I]{bacon01}. The main goals of the project are to study
the intrinsic shapes, velocity and metallicity distributions of the sample
galaxies, and to gain insight into the relation between their stellar/gaseous
kinematics and stellar populations. Details of the definition and properties of
the survey can be found in \citet[hereafter Paper II]{tim02}. Previous papers of
the series concentrated on the presentation and characterization of the stellar
and ionised-gas kinematics and absorption line-strength indices of the sample of
48 elliptical and lenticular galaxies \citep[hereafter Papers III, V and
VI]{emsellem04,sarzi05,kuntschner06}. Here, we present the combined analysis of
the stellar and ionised-gas distribution and kinematics of the 24 Sa galaxies
in the representative sample. In addition, we investigate the importance of
ongoing star formation in the sample, and its relation to the dynamical state
of the galaxies. For a more comprehensive study of an individual case in our
sample (i.e., NGC\,5448) we refer the reader to \citet{fathi05}. \citet{ganda05}
discuss similar \sauron\ observations of later-type spiral galaxies. The data
and maps presented here will be made available via the \sauron\ website
(http://www.strw.leidenuniv.nl/sauron).

The paper is structured as follows. Section~\ref{sec:observations} summarises
the observations, instrumental setup and data reduction.
Section~\ref{sec:analysis} describes the tools employed to extract the stellar
and ionised-gas kinematics. Section~\ref{sec:liter_comp} shows the comparison of
our results with aperture measurements in the literature.
Section~\ref{sec:results} is devoted to the presentation of the stellar and
ionised-gas maps for each galaxy, as well as the description of the main
features. Global and circumnuclear star formation is discussed in
Section~\ref{sec:sfr}, and the results are summarised in
Section~\ref{sec:conclusions}. A more detailed description of the structures in
the individual galaxies is presented in Appendix~\ref{sec:galaxies_notes}.

\section{Observations and Data Reduction}
\label{sec:observations}

\subsection{Observing runs and instrumental setup}
\label{sec:runs}
Observations of the 24 Sa galaxies were carried out using the integral-field
spectrograph \sauron\ attached to the 4.2-m William Herschel Telescope (WHT) of
the Observatorio del Roque de los Muchachos at La Palma, Spain. For each of the
24 galaxies, Table~\ref{tab:expo} lists the run in which it was observed and the
exposure time for each individual pointing. More details on the observing
conditions in each run can be found in Paper III.

We used the low spatial resolution mode of \sauron, providing a
$33\arcsec\times41\arcsec$ field-of-view (FoV). The spatial sampling is
determined by an array of $0\farcs94\times0\farcs94$ square lenses. This setup
produces 1431 spectra per pointing over the \sauron\ FoV. Additionally, $146$
lenses provide simultaneous sky spectra $1\farcm9$ away from the main field. 
\sauron\ delivers a spectral resolution of 4.2~\AA\ (FWHM) and covers the narrow 
spectral range 4800-5380~\AA. This wavelength range includes a number of important
stellar absorption lines (e.g. \hbeta, Fe5015, Mg$b$, Fe5270) and also
potential emission lines (\hbeta$\lambda4861$, \oiii$\lambda\lambda$4959,5007,
\ni$\lambda\lambda$5198,5200).

\begin{figure*}
\begin{center}
  \includegraphics[scale=0.8]{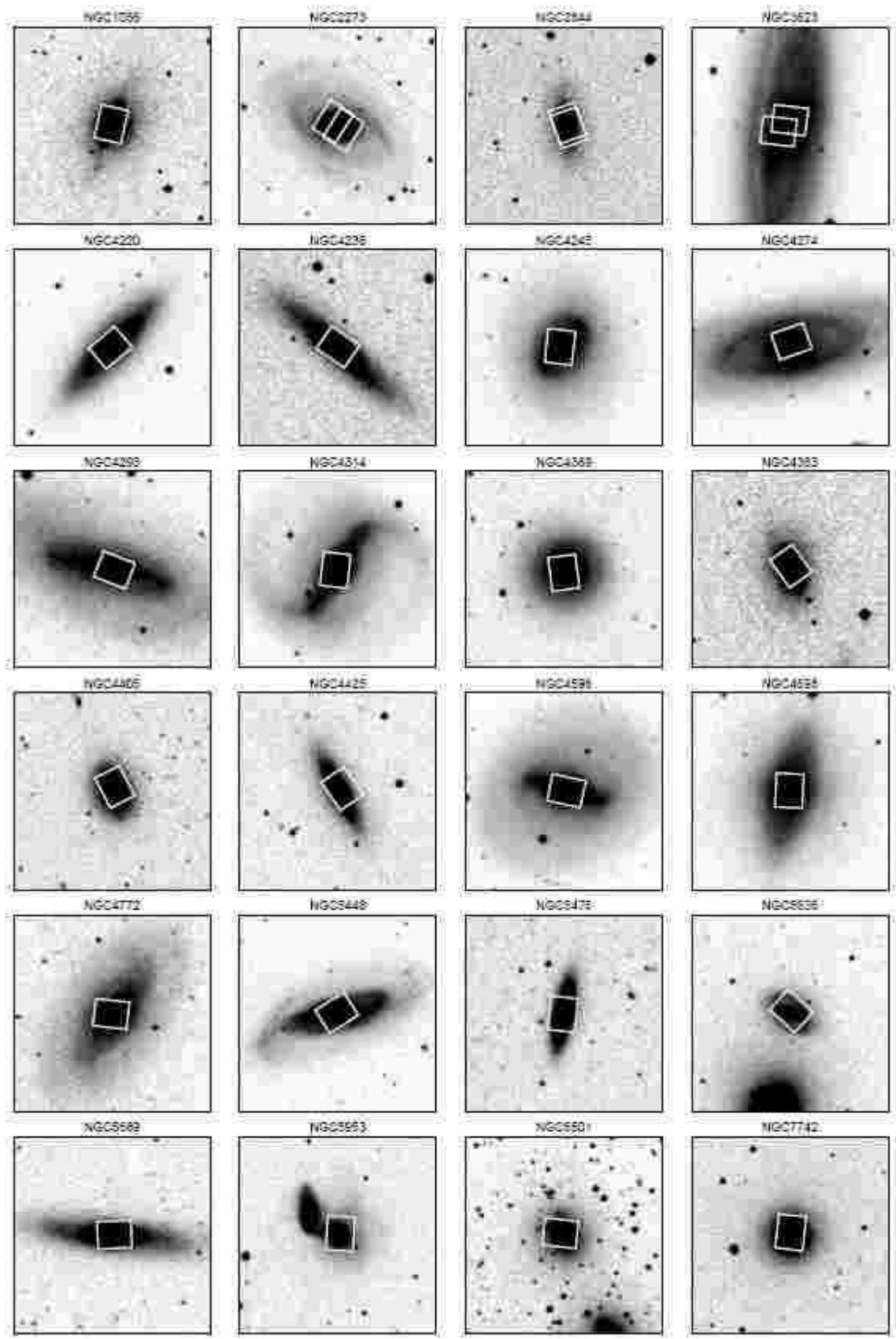}
\end{center}
\caption{$R$-band Digital Sky Survey images of the 24 Sa spiral galaxy bulges in
the \sauron\ representative sample. The size of each image is
4\arcmin$\times$4\arcmin, and the orientation is such that north is up 
and east is left. Overlaid on each image is the approximate field-of-view of the
\sauron\ pointings obtained for the object.}
\label{fig:dss}
\end{figure*}

For each galaxy, four overlapping exposures of 1800\,s were typically obtained.
An offset of a few arcseconds, corresponding to a small, non-integer number of
spatial elements, was introduced between exposures to avoid systematic errors
due to e.g. bad CCD regions. While for most of the galaxies a single pointing
was sufficient to cover the bulge dominated region, in three cases (NGC\,2273,
NGC\,2844, NGC\,3623) two pointings were required. Footprints of these pointings
are shown overlaid on $R$-band Digital Sky Survey images in
Figure~\ref{fig:dss}.

\begin{table}
\begin{center}
\caption{Details of the exposures of the \sauron\ Sa spiral bulges.}
\label{tab:expo}
\begin{tabular}{lccclccc}
\hline
\multicolumn{4}{c}{Sa galaxies: Field} &  \multicolumn{4}{c}
{Sa galaxies: Cluster}\\
\hline
NGC &Run &~$\#$ &T$_{\rm exp}$ & NGC &Run &~$\#$ &T$_{\rm exp}$\\
~(1) &(2) &(3) &(4)&  ~(1) &(2) &(3) &(4)\\
\hline
1056 & 6 & 1 & 4x1800 &  3623 & 3 & 1 & 4x1800 \\
2273 & 5 & 1 & 4x1800 &       & 3 & 2 & 4x1800 \\
     & 5 & 2 & 4x1800 &  4235 & 7 & 1 & 4x1800 \\
2844 & 6 & 1 & 4x1800 &  4245 & 7 & 1 & 4x1800 \\
     & 6 & 2 & 2x1800 &  4274 & 7 & 1 & 4x1800 \\
4220 & 8 & 1 & 3x1800 &  4293 & 8 & 1 & 3x1800 \\
4369 & 7 & 1 & 4x1800 &  4314 & 7 & 1 & 4x1800 \\
5448 & 7 & 1 & 4x1800 &  4383 & 7 & 1 & 4x1800 \\
5475 & 7 & 1 & 6x1800 &  4405 & 7 & 1 & 4x1800 \\
5636 & 7 & 1 & 5x1800 &  4425 & 7 & 1 & 4x1800 \\
5689 & 5 & 1 & 5x1800 &  4596 & 7 & 1 & 4x1800 \\
5953 & 5 & 1 & 4x1800 &  4698 & 7 & 1 & 4x1800 \\
6501 & 3 & 1 & 4x1800 &  4772 & 3 & 1 & 4x1800 \\
7742 & 2 & 1 & 4x1800 &       &   &   &        \\
\hline
\end{tabular}
\\
Notes:
(1)~NGC number.
(2)~Run number (see Table~1 in Paper III).
(3)~Pointing number.
(4)~Exposure time, in sec.
\end{center}
\end{table}

\begin{table*}
\caption{Characteristics of the \sauron\ Sa galaxies.}
\label{tab:allgal}
\begin{center}
{\tabcolsep=3.5pt
\begin{tabular}{llcccccccccccccccc}
\hline
NGC & Type & V$_{\rm syst}$ &  Source & Seeing   & $\theta_s$  & 
$\Delta\phi_{phot}$ &  $\Delta\phi_{kin}$  &    $\rm{log}~F_{\hbeta}$   & 
$\rm{A}_{gal}$  &  $\rm{A}_{int}$  & $\rm{log}~L_{\halpha}$  & $\rm{log}~SFR$ 
  &  $\rm{log}~M_{\hii}$   \\
  ~ &   ~  &  [\kms]        &         & [arcsec] &   [deg]     &   [deg]      &     [deg]      &   [erg s$^{-1}$ cm$^{-2}$] &	[mag]  &  [mag]  &     [erg s$^{-1}$]    & [\Msun\ yr$^{-1}$]  &       [\Msun]	       \\	  
 ~(1) & (2) & (3) & (4) & (5) & (6)   &  (7)  & (8)  & (9)  & (10) & (11) & (12) & (13)  & (14) \\
\hline
1056 & Sa:         & 1564  & WHT     & 2.0  &   96 & 	4 &   6 & -12.92 & 0.493 & 0.167 & 40.51  & -0.59  & 5.88 \\
2273 & SBa(r):     & 1852  & F606W   & 1.7  &  206 & 	4 &  12 & -13.09 & 0.235 & 0.122 & 40.37  & -0.73  & 5.74 \\
2844 & Sa(r):      & 1510  & WHT     & 1.5  &  243 & 	1 &   5 & -13.25 & 0.062 & 0.213 & 40.01  & -1.10  & 5.37 \\
3623 & SABa(rs)    &  841  & F814W   & 1.4  &    0 & 	8 &  14 & -13.13 & 0.083 & 0.357 & 39.68  & -1.42  & 5.05 \\
4220 & S0$^+$(r)   &  946  & F814W   & 2.3  &  313 &   15 &   9 & -13.34 & 0.059 &   -   & 39.42  & -1.68  & 4.79 \\
4235 & Sa(s)sp     & 2295  & F606W   & 1.2  &  208 & 	2 &   5 & -13.63 & 0.062 & 0.357 & 40.05  & -1.05  & 5.42 \\
4245 & SB0/a(r)    & 904   & F606W   & 1.5  &  90  &   18 &   4 & -13.39 & 0.069 & 0.053 & 39.36  & -1.75  & 4.73 \\
4274 & (R)SBab(r)  & 930   & F555W   & 1.7  &  154 & 	5 &  21 & -13.22 & 0.074 & 0.319 & 39.66  & -1.45  & 5.03 \\
4293 & (R)SB0/a(s) & 948   & F606W   & 2.8  &  195 &   30 &  38 & -13.61 & 0.130 & 0.122 & 39.23  & -1.87  & 4.60 \\
4314 & SBa(rs)     & 1001  & F814W   & 1.6  &  90  &   24 &   5 & -12.83 & 0.083 & 0.015 & 40.00  & -1.10  & 5.37 \\
4369 & (R)Sa(rs)   & 1043  & WHT     & 1.4  &  76  &   29 &  12 & -12.41 & 0.085 & 0.015 & 40.45  & -0.65  & 5.82 \\
4383 & Sa pec      & 1724  & F606W   & 0.9  &  228 & 	1 &   5 & -12.22 & 0.078 & 0.198 & 41.15  &  0.04  & 6.51 \\
4405 & S0/a(rs)    & 1750  & F606W   & 1.8  &  236 & 	5 &   2 & -13.14 & 0.080 & 0.068 & 40.19  & -0.91  & 5.56 \\
4425 & SB0$^+$:sp  & 1919  & WHT     & 1.1  &  229 & 	3 &   - &    -   &   -   &   -   &   -    &   -    &  -   \\
4596 & SB0$^+$(r)  & 1910  & F606W   & 2.2  &  6   &   16 &   8 & -13.72 & 0.074 &   -   & 39.66  & -1.44  & 5.03 \\
4698 & Sab(s)      & 1035  & F606W   & 1.1  &  86  & 	3 &   7 & -13.17 & 0.086 & 0.213 & 39.51  & -1.59  & 4.88 \\
4772 & Sa(s)       & 1068  & F606W   & 1.6  &  0   &   34 &  30 & -13.89 & 0.090 & 0.175 & 39.06  & -2.05  & 4.43 \\
5448 & (R)SABa(r)  & 2043  & F606W   & 1.1  &  141 &   13 &   1 & -13.48 & 0.049 & 0.190 & 40.03  & -1.08  & 5.39 \\
5475 & Sa sp       & 1667  & F814W   & 1.3  &  90  & 	2 &  57 & -13.95 & 0.039 & 0.289 & 39.40  & -1.70  & 4.77 \\
5636 & SAB(r)0+    & 1665  & WHT     & 1.5  &  216 &   15 &  10 & -13.90 & 0.109 &   -   & 39.37  & -1.73  & 4.74 \\
5689 & SB0$^0$(s)  & 2225  & F814W   & 1.9  &  171 & 	2 &   9 & -14.01 & 0.118 & 0.160 & 39.58  & -1.52  & 4.95 \\
5953 & Sa:pec      & 2010  & F606W   & 1.5  &  87  &   13 &   4 & -12.60 & 0.162 & 0.076 & 40.88  & -0.22  & 6.25 \\
6501 & S0$^+$      & 2967  & WHT     & 2.5  &  0   & 	6 &   - &    -   &   -   &   -   &   -    &   -    &  -   \\
7742 & Sb(r)       & 1678  & F555W   & 1.6  &  270 &   30 &   1 & -12.50 & 0.182 & 0.023 & 40.82  & -0.28  & 6.19 \\
\hline
\end{tabular}}
\end{center}
\begin{minipage}{17.8cm}
Notes:
(1)~NGC number. (2)~Hubble type from NED (http://nedwww.ipac.caltech.edu/). 
(3)~Estimate of the heliocentric systemic velocity, in \kms, corrected for the 
barycentric motion.
(4)~Source for the seeing determination (see Sect.~\ref{sec:imaging}). 
(5)~Seeing, full width at half maximum in arcsec. 
(6)~Position angle, in degrees, of the vertical (upward) axis in the maps shown 
in Figs.~\ref{fig:N1056}-\ref{fig:N7742}. (7)~Misalignment between the 
photometric and stellar kinematical major axis in degrees 
($\Delta\phi_{phot}=\vert\phi_{\rm phot}-\phi_{\rm star}\vert$). 
(8)~Misalignment between the major axis of the stellar and ionised-gas 
kinematics ($\Delta\phi_{kin}=\vert\phi_{\rm star}-\phi_{\rm gas}\vert$). 
The mean uncertainty in the measured misalignments in columns 7 and 8 is 
6 and 7 degrees, respectively (see Section~\ref{subsec:misalignments}). 
(9)~Observed \hbeta\ flux in the \sauron\ FoV.
(10)~$V$-band galactic extinction from \citet{schlegel98} as listed in NED.
(11)~$V$-band dust extinction due to inclination from \citet{bottinelli95}. The 
original values in $B$-band, listed in Hyperleda, were converted to $V$-band 
assuming the attenuation law of \cite{cardelli89}.
(12)~Estimated extinction corrected \halpha\ luminosity. 
(13)~Total star formation rate within the \sauron\ FoV. 
(14)~Mass of the ionised gas within the \sauron\ FoV.
(See Section~\ref{sec:sfr} for details).\looseness-2
\end{minipage}
\end{table*}

\subsection{Data reduction}
\label{sec:reduction}
We followed the procedures described in Paper I and III for the extraction, 
reduction, and calibration of the data, using the specifically designed \Xsauron\
software developed at CRAL. For each galaxy, the sky level was measured using
the dedicated sky lenses and subtracted from the target spectra. Arc lamp
exposures were taken before and after each target frame for wavelength
calibration. Tungsten lamp exposures were also taken at the beginning and end of
each night in order to build the mask necessary to extract the data from the CCD
frames. Flux and Lick absorption line standard stars were observed during each
observing run for calibration purposes. Since the flux standards and the
galaxies were not always observed under the same conditions, the fluxes quoted 
in this paper are approximate. Differences in the flux calibration between 
observing runs are, however, smaller than 6\% (see Paper VI). The individually 
extracted and flux calibrated datacubes were finally merged by truncating the 
wavelength domain to a common range and spatially resampling the spectra to a 
common squared grid. The dithering of individual exposures enabled us to sample 
the merged datacube onto $0\farcs8\times0\farcs8$ pixels.

During the eighth observing run, the grating of the instrument was replaced by a
Volume Phase Holographic grating (VPH). This change gave an increase in 
sensitivity of the instrument. Only two galaxies in our sample are affected by
this change (NGC\,4220, NGC\,4293). A specifically designed mask model (see
Paper I for details) was produced for the extraction of the data in this run.
All the procedures described above are valid for both configurations.

In order to ensure the measurement of reliable stellar kinematics, we spatially
binned our final datacubes using the Voronoi 2D binning algorithm of
\citet{michele03}, creating compact bins with a minimum signal-to-noise ratio
($S/N$) of $\sim60$ per spectral resolution element. Most spectra in the central
regions, however, have $S/N$ in excess of $60$, and so remain un-binned.

\subsection{HST and ground-based imaging}
\label{sec:imaging}
In addition to the \sauron\ data, we retrieved from the {\it Hubble Space
Telescope} ({\it HST}) archive\footnote[1]{Based on observations made with the
NASA/ESA Hubble Space Telescope, obtained from the ESO/ST-ECF Science Archive
Facility.} the available WFPC2 imaging closest to the F555W filter. The origin 
of the images is rather heterogeneous, so the final set of images used in this
paper comes from observations made in several photometric filters. Since no
{\it HST}/WFPC2 image was available for NGC\,4220, ACS/F814W imaging was used 
instead. In the cases with no {\it HST} or with poor signal-to-noise imaging 
(e.g., {\it HST}/WFPC2/F300W) we used ground-based photometry, in the F555W and 
F814W filters, from the 1.3m McGraw-Hill telescope at the MDM observatory on 
Kitt Peak. The MDM images are part of a complete photometric survey of the 
\sauron\ galaxies and will be described elsewhere.

The use of this photometric dataset is two-fold. First, it allows us to generate
unsharp-masked images, to highlight any small scale structures that might be
present in the galaxies. Second, we used the HST images to accurately determine
the point spread function (PSF) of each merged exposure, by comparing them to
the reconstructed \sauron\ intensity distribution. We followed the procedure
outlined in Paper III to derive the PSF values. Where HST imaging is not
available, we quote the value from the seeing monitor during the observations.
Table~\ref{tab:allgal} summarises the HST imaging used (i.e., in the available
filter closest to the V-band), and lists the seeing estimate for each galaxy.
Cases with no HST imaging are indicated by WHT.\looseness-2

\section{Analysis and Methods}
\label{sec:analysis}

\subsection{Stellar kinematics}
\label{subsec:stekin}

We measured the stellar kinematics of the 24 Sa spiral bulges using the
penalized pixel-fitting (pPXF) method of \citet{capem04}. We made use of the
library of single age, single metallicity population (SSP) models from
\citet{vazdekis99} as stellar templates. We selected a representative set
of 48 SSP models evenly sampling a wide range in age and metallicity (1.00 $\le$
Age $\le$ 17.38 Gyr, -1.68 $\le$ [Fe/H] $\le$+0.20). This set of templates
differs from the one used in Paper III in that we expanded the range of
metallicities to include more metal-poor templates. This was necessary to be 
able to accurately match the main absorption features in our galaxies. Exclusion 
of these models resulted in a poor match of the overall spectrum and therefore a
bias in the measured parameters.\looseness-2

A non-negative linear combination of SSP models, convolved with a Gauss-Hermite
series \citep{vdm93,gerhard93}, was fitted to each individual spectrum. The
best-fitting parameters were determined by chi-squared minimization in pixel
space. In the wavelength range covered by \sauron, a few potential emission
lines were masked during the fitting procedure (\hbeta, \oiii, \ni).
Additionally, a low-order Legendre polynomial was included in the fit to account
for small differences in the continuum shape between the galaxy spectra and the
input library of synthetic models.\looseness-1

Our sample of spiral galaxies displays stellar velocity dispersions that are, in
general, much lower than those found in the ellipticals and lenticulars in
Paper III. Given the instrumental sampling of \sauron\ (60 \kms\ per pixel),
some bias in the determination of the stellar velocity dispersion might be
expected for values below $\sim$120 \kms. This bias can me minimised by choosing
an adequate penalisation factor ($\lambda$) in pPXF. We carried out simulations 
to determine the best factor for our dataset, and a value of $\lambda=0.7$ was 
adopted, as in Paper III. The level of uncertainty of the velocity dispersion 
$\sigma$, for this penalization factor, was extensively tested in Paper III by 
means of Monte-Carlo simulations. As shown in Figure~2 of that paper, for a 
spectrum with a $S/N$ of $\sim60$ and an intrinsic $\sigma$ as low as 50 \kms, 
the measured velocity dispersion differs from the intrinsic one by, at most, 
$\sim$10 \kms\ even for extreme $h_3$ and $h_4$ values, which is within the 
estimated errors. In practice only the outer bins in our maps (i.e. the ones 
with the lowest velocity dispersions) are affected by this bias.

\begin{figure}
\begin{center}
   \includegraphics[width=0.99\linewidth]{comparison_skin.eps2}
   \includegraphics[width=0.99\linewidth]{comparison_Ho.eps2}
\end{center}
\caption{Comparison of kinematic parameters with the literature. Top: Central
aperture measurements of the stellar velocity dispersion of the \sauron\ Sa
spiral galaxies and values taken from Hyperleda. Bottom: Central aperture
measurements of the \oiii\ ionised-gas FWHM of the \sauron\ Sa spiral galaxies
and \nii\ values taken from the Palomar spectroscopic survey \citep{ho95,ho97}.
Solid line represents a least-squares fit to the data points. The dotted line
marks the 1:1 relation.}
\label{fig:comparison}
\end{figure}

\subsection{Gas kinematics}
\label{subsec:gaskin}

Typical spectra from our sample of galaxies contain significant amounts of
nebular emission in the \sauron\ wavelength range. However, stellar absorption
features are still dominant overall, and therefore the measurement of the gas 
distribution and kinematics requires a careful separation of the line emission 
from the stellar absorption. For this, we follow the procedure described in 
Paper V. Briefly, this method consists of searching iteratively for
the emission-line velocities and velocity dispersions, while linearly solving at
each step for both their amplitudes and the optimal combination of the stellar 
templates over the full \sauron\ wavelength range. No masking of the regions
affected by emission is thus required. The stellar kinematics is held fixed 
during the fitting process. A low-order Legendre polynomial (typically of order 
6) is included in the fit to account for small differences between the galaxy 
spectra and the input library of synthetic models. We refer the reader to Paper 
V for a more detailed description of the method, limitations, and sensitivity 
limits.

In general, the amount of emission in both \hbeta\ and \oiii\ is important in 
our sample. In some galaxies we are also able to detect significant emission 
from the \ni\ doublet, but in most cases this is very weak and it is difficult 
to establish its kinematics. As observations of the \halpha\ and \nii\ lines 
(e.g., \citealt{sttht98}), and the \ni\ and \nii\ emission lines 
(e.g., \citealt{ge96}) display similar kinematics, we do not expect to find 
significant differences between the kinematics of \hbeta\ and \ni. Given the 
small amount of information the \ni\ maps add to our results, we do not show 
them here. They will be made available with the public data release.\looseness-2

As the ionised gas is collisional and dissipative, the emission-line maps can
exhibit complicated morphologies and kinematics \citep{plana98,amp05,chemin05,
chemin06}. In practice this could translate into complex emission-line profiles
due to the superposition of distinct gas clouds. We carefully inspected the
emission line profiles in our data to search for asymmetric profiles, but
we could not find any significant deviation from a pure Gaussian in the majority
of our galaxies. Only a few cases revealed complex profiles in specific
regions close to their centre, possibly related to their AGN nature (e.g.,
NGC\,2273). The lack of such complex profiles in the rest of our sample may be, 
at least in part, due to \sauron's limited spectral resolution. We therefore 
fitted a single Gaussian profile to each line in our wavelength range. We fixed 
the amplitude ratio of the \oiii\ emission-line doublet to 1:3 \citep{osterbrock},
and to 0.7:2 for the \ni\ lines\footnote{Ratio estimated using the Mappings Ic
software with an electron temperature of 10$^{4}$ K and electron densities from
0.1 cm$^{-3 }$ to 1000 cm$^{-3 }$ \citep{fbsp97}}. As gas clouds may be ionised
by different mechanisms, it is possible that in certain regions of the galaxies
the ionisation of \hbeta\ is more efficient than \oiii\ and viceversa. Given
that in general our sample galaxies display very strong emission, we looked for
differences in the kinematics by fitting these two emission lines independently.
Based on the simulations performed in Paper V, we considered true detection of
emission when the ratio of the amplitude of the emission line to the surrounding
noise ($A/N$, see Paper V) is larger than 5, 4, and 4 for \hbeta, \oiii, and 
\ni\ respectively.

\begin{figure*}
\begin{center}
   \includegraphics[angle=0,width=0.99\linewidth]{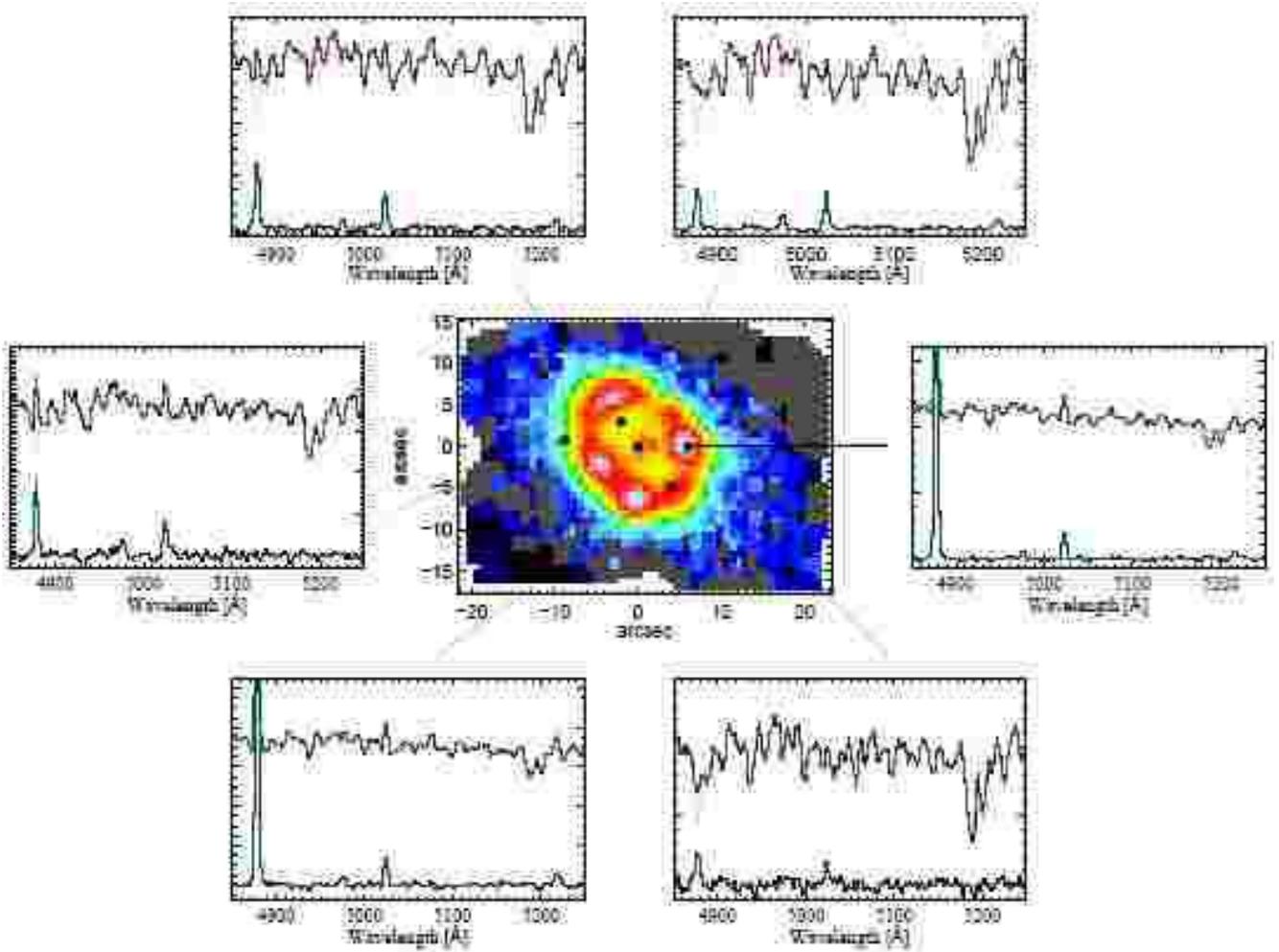}
\end{center}
\caption{Fits of the \sauron\ spectra of NGC\,4314. The inset corresponds to the
\hbeta\ flux map (see also Fig.~\ref{fig:N4314}). Data bins with A/N values
below 5 (see~\ref{subsec:gaskin}) are displayed in the different maps as dark
grey. Six spectra and their corresponding fits at different locations are shown
around the inset image. In each panel the black line shows the galaxy spectrum,
the red line is the best fitting stellar template, and the blue line is the best
fit when emission lines are added. Below each spectrum the differences between
the galaxy spectrum and the best stellar template are shown in black. The
emission-line fits are overplotted in green. The spectra have been arbitrarily
shifted in the vertical direction for presentation purposes.}
\label{fig:optemp}
\end{figure*}

\section{Comparison with the Literature}
\label{sec:liter_comp}
As demonstrated in Papers II, III and V for E and S0 galaxies, the methods used
to derive the stellar and ionised-gas kinematics produce comparable results
with those published in the literature. Here, we report similar comparisons
to assess the robustness of these methods for our sample of Sa bulges, where
emission is much more prominent and stellar velocity dispersions are smaller.

Literature measurements for our sample of Sa bulges are relatively scarce and
come from heterogeneous sources, which often also implies different aperture
sizes and slit position angles. We used the compilation in the on-line catalogue
Hyperleda\footnote{http://leda.univ-lyon1.fr/} \citep{ps96} for the comparison
of the stellar kinematics, and the Palomar spectroscopic survey
\citep{ho95,ho97} for the line-width of the ionised gas. We measured the stellar
velocity dispersions and FWHM of the \oiii\ emission lines from a standard
central aperture of $2\arcsec \times 4\arcsec$ extracted from our \sauron\ maps.
This aperture size was chosen to match the apertures in the Palomar survey. For
the stellar velocity dispersions, apertures were aligned along the axis reported
in the individual references (i.e. \citealt{terlevich90, dinella95, rampazzo95,
bernardi02, fbpv03}). We only considered literature values with reported errors
below 10 \kms\ (i.e., the typical error estimate of our measurements). For the
ionised-gas measurements, we used the position angle of the Palomar long-slit
observations. We excluded from the comparison galaxies with strong non-Gaussian
emission-line profiles (i.e., AGN-like), as the single Gaussian used in our
method is not an accurate representation of those line profiles. Uncertainties 
in the \nii\ FWHM in the Palomar survey were typically 10\%, except in 
NGC\,4596 for which the reported uncertainty is 40\%.

On the top panel of Figure~\ref{fig:comparison} we show the result of the
comparison for the stellar velocity dispersions. A linear regression was
fitted to the combined datasets. We made use of the {\it FITEXY}\footnote{Based
on a similar routine by \citet{press92}} routine taken from the IDL
Astro-Library \citep{landsman93} to fit a linear relation of the form
$\mathrm{y}=\alpha+\beta~(\mathrm{x-x_{0}})$, where $\mathrm{y}=\sigma_{\rm{SAURON}}$
and $\mathrm{x}=\sigma_{\rm{LIT.}}$. The value of $x_{0}$ was chosen to be 150 \kms\
to minimise the uncertainty in the fitted intercept. The resultant fit provides
a slope of $1.01\pm0.10$ and an intercept of $150.40\pm3.53$ \kms. Our
measurements are thus in good agreement with those in the literature. On the
bottom panel we plot the results for the ionised-gas FWHMs. A similar linear
regression fit gives a slope of $0.98\pm 0.14$ and intercept of $172.43\pm
13.36$ \kms. While a correlation between the \oiii\ FWHM and \nii\ FWHM may be
expected if these lines are produced in low-density reservoirs \citep{ho97b},
given the strong dependence of the line widths on the critical densities it is
possible that this assumption does not hold for all galaxies in our sample.
Considering this result and all the possible systematic effects at play in this
comparison, our measurements seem to be in good agreement with those of
\citet{ho97}. 

Additionally, we tested the reliability of our measurements by checking the
quality of the optimal template fits to the spectra in our galaxies. In
Figure~\ref{fig:optemp} we show these fits for one of the most critical cases in
our sample: NGC\,4314 \citep{benedict92}. We chose to illustrate this example
because of the wide range of emission-line fluxes in the \hbeta\ map relative 
to the underlying stellar continuum. In this galaxy, most of the \hbeta\
emission is confined to a ring of radius $\sim$10\arcsec. Outside the ring the
amount of \hbeta\ flux decreases by more than a factor 100. The different panels
in this figure demonstrate that our method is able to produce very good fits to
the spectra in several distinct regions of the galaxy, and that our results are
not biased because of an inadequate matching of the spectra. Similar agreement
is found for the rest of the sample of galaxies.

\section{Observed Stellar \& Gas Kinematics}
\label{sec:results}
Figures~\ref{fig:N1056}-\ref{fig:N7742} display maps of the absorption and
emission-line distribution and kinematics of the 24 Sa galaxies in our sample.
The maps are displayed according to increasing NGC number. In each case, we show
({\it first row}) an unsharp-masked image of the galaxy from either HST or
ground-based MDM data (see Section~\ref{sec:imaging}). The {\it second row}
displays the total intensity reconstructed from the \sauron\ spectra (in
mag/arcsec$^2$ with an arbitrary zero point), the mean stellar radial velocity
$V$, and velocity dispersion $\sigma$ (both in \kms). The {\it third} and {\it
fourth} row presents the flux (in logarithmic scale), mean velocity and velocity
dispersion (both in \kms) of the \hbeta\ and \oiii\ emission lines,
respectively. The {\it fifth} row shows the \oiii/\hbeta\ ratio map (in
logarithmic scale), as well as the stellar Gauss--Hermite velocity moments $h_3$
and $h_4$. The maps of the ionised gas were constructed according to the $A/N$
ratios set in \S\ref{subsec:gaskin}. Data bins with $A/N$ values below these
thresholds are displayed in dark grey. Mean stellar radial velocities are with
respect to estimated heliocentric systemic velocities, the values of which are
provided (corrected for the barycentric motion) in Table~\ref{tab:allgal}. The
same heliocentric systemic velocity is assumed in the emission-line velocity
maps.

Detailed descriptions of the different maps are collected in
Appendix~\ref{sec:galaxies_notes}. Here, we concentrate on an overview of
the general trends and results observed.

\subsection{Stellar kinematics}
\label{subsec:stars}
Inspection of the different figures reveals a wide range of kinematic
structures. The presence of dust in our galaxies can have, in many cases, a
strong impact on the measured stellar kinematics, as it affects the amount of
light we receive along the line-of-sight. Despite the dust, and as expected for
Sa galaxies, the maps show clear stellar rotation even at low galaxy
inclinations (e.g., NGC\,7742). Kinematically decoupled components occur
frequently in the inner regions (NGC\,1056, NGC\,2273, NGC\,3623, NGC\,4235,
NGC\,4245, NGC\,4274, NGC\,4596, NGC\,5448, NGC\,5689, NGC\,7742). These
structures are usually detected in the stellar velocity maps alone (as a sudden
change in velocity, or pinching of the isovelocity contours close to the
centre), but often they are also associated with a drop in the velocity
dispersion and anti-correlated $h_3$ values with respect to the stellar
velocities. Moreover, they are generally related to features in the photometry,
such as bars or central dust discs and rings. These structures are in nearly
all cases co-rotating and aligned with the main body of the galaxy. The
flattening of the isovelocities of these components suggests that these might
be inner discs or rings. In addition, there are a few galaxies in the sample 
(NGC\,4698 and NGC\,5953) that have kinematically decoupled components which are 
misaligned with respect to the major axis of the galaxy.

Velocity dispersion minima (i.e., `velocity dispersion drops') are common in our
sample. We observe distinct dispersion drops in 11 out of 24 galaxies (46\%).
This is probably a lower limit considering the medium spatial resolution of our
\sauron\ data. Only 4 of these 11 cases are known to have an active nucleus.
Recent long-slit studies \citep{cb04} revealed a velocity dispersion drop
frequency of about 40\% in a sample of 30 nearly edge-on S0-Sbc galaxies, which
is consistent with our findings. The first observed cases of central velocity
dispersion minima  date back to the late 80s and early 90s (e.g.,
\citealt{bottema89, bottema93, fisher97}). However, this subject has started to
gain attention in the last years (e.g., \citealt{emsellem01,marquez03}). Despite
the growth of observational examples, the nature of these drops is not well
understood. First theoretical predictions for the presence of these velocity
dispersion minima \citep{binney80} suggested that they were the natural
consequence of inner regions of galaxies following an r$^{1/4}$ light profile
\citep{devauc48}. Later on this prediction was extended to more general
r$^{1/n}$ profiles \citep{sersic} by \citet{cl97}. In practice, these drops are
not common in elliptical galaxies (Paper III). The objects in our sample are Sa
galaxies with generally large bulges dominating the light in the inner regions.
If bulges were scaled-down versions of ellipticals, then one would not expect
velocity dispersion drops in Sa galaxies. Therefore, the observations lead to
the picture that Sa bulges may have disk properties as well \citep{kormendy93}.
The latest N-body simulations associate the origin of these dispersion minima to
kinematically cold components (i.e. discs), formed by gas inflow towards the
central regions of the galaxy and subsequent star formation \citep{wozniak03}.
However, the importance of dissipative processes \citep{ba05}, and the role of
bars \citep{hs94,fb95} is still unclear in this context.

A large fraction of the galaxies with sigma-drops do indeed show the presence of
star formation at the same locations (as seen from the low \oiii/\hbeta\ 
ratio). The presence of star formation in the inner regions of galaxies, 
however, does not always translate into a stellar velocity dispersion drop. 
Given the typical gas consumption time scale in normal disk galaxies ($\sim$ 1 
Gyr), it is possible that many of the galaxies displaying these drops, but not 
young star formation, have simply used up all the available gas to form 
stars. The detection of the drop can also be more difficult due to the 
inclination of the galaxy, as in face-on configurations the contribution of the 
dynamically cold stars to the line-of-sight is much smaller than that of the 
surrounding bulge. Examples of this effect are found in NGC\,4274 and NGC\,4314. 
A more in-depth discussion of the link between the velocity dispersion drops and 
their underlying stellar populations will be the subject of a future paper in 
this series.\looseness-2

%

Some objects in the sample (NGC\,2273, NGC\,3623, NGC\,4220, NGC\,4245,
NGC\,4293, NGC\,4369, NGC\,4596, NGC\,5448, NGC\,5636) display misaligned
photometric and kinematic axes, indicative of a non-axisymmetric structure
(i.e., a bar). These misalignments are more easily detected at low galaxy
inclinations, but it is still possible to recognise the existence of bars by
means of their kinematic signatures even for edge-on configurations. Following
the pioneering work of \citet{kuijken95}, recent N-body simulations of barred
galaxies \citep{ba05} have suggested several bar diagnostics making use of the
parameters of the Gauss-Hermite series ($V$, $\sigma$, $h_3$, $h_4$). In our
sample we find at least five highly-inclined galaxies (NGC\,3623, NGC\,4235,
NGC\,4274, NGC\,5448, NGC\,5689) that reveal some of the main kinematic
signatures shown in the simulations: double-hump rotation curve, broad velocity
dispersion profile with a plateau at moderate radii, and $h_3-V$ correlation
over the projected bar length. In addition to these bar diagnostics, other
features such as boxy isophotes or cylindrical rotation hint at the presence of
a bar in NGC\,4220 and NGC\,4425. Nuclear star forming rings are also often
interpreted in the context of bar driven evolution. We return to this point in
Section~\ref{sec:sfr}.

\subsection{Ionised-gas morphology and kinematics}
\label{subsec:gas}
As expected, nearly all the galaxies in our sample display significant amounts
of ionised-gas emission. There are only two cases where the detection of ionised
gas is marginal or absent (NGC\,4425, NGC\,6501). Similarly to the stellar
kinematics, we can also correlate the presence and spatial morphology of the
dust with the distribution seen in the \sauron\ ionised-gas maps. Dust features
in the unsharp-masked images are usually associated with the presence of
ionised-gas structures. However, the inverse is not always true. This may be due
to the fact that dust is much more difficult to detect in face-on systems or is
simply not there. The connection between ionised gas and dust is more evident in
the \hbeta\ than in the \oiii\ emission, which overall also tends to be stronger
than \oiii. This difference is particularly apparent in the \oiii/\hbeta\ maps.
A large fraction of the objects in our sample display very low \oiii/\hbeta\
ratios, suggesting currently ongoing star formation. The morphology of star
formation regions is varied. There are six cases in our sample where star
formation occurs in ring-like structures (see Sect.~\ref{sec:sfr}). In some
other galaxies, however, the star-forming regions spread across the main body of
the galaxy (e.g., NGC\,1056), align with strong dust lanes (e.g., NGC\,4220) or
have a more amorphous morphology (e.g., NGC\,4369, NGC\,4405). In most of the
cases, these low \oiii/\hbeta\ regions have gas velocity dispersion values below
$\sim$50 \kms. This suggests that there is a large amount of cold gas present,
from which stars have recently formed.\looseness-2

The velocity maps of the ionised gas in our sample are consistent with gas 
motions in a disc co-rotating with the stars. In these cases the rotational 
velocities exhibit rotation amplitudes larger than those of the stars, as 
expected from the dissipative nature of the gas (and the corresponding smaller 
asymmetric drift). A few cases depart from this simple description: NGC\,4772 
and NGC\,5953, where the gas in certain regions rotates almost perpendicularly 
to the stars; NGC\,7742, where the gas counter-rotates with respect to the 
stars; NGC4383, which displays mean velocities confined in a jet-like structure; 
NGC4369, that shows no well defined rotation axis, possibly due to the patchy 
nature of the gas distribution, and NGC\,3623, where the gas traces spiral arm 
structure. Our maps also display differences in the \hbeta\ and \oiii\ mean 
velocities and velocity dispersions. This is particularly apparent in the inner 
arcseconds (e.g., NGC\,4698, NGC\,5475), and sometimes also outside the nuclear 
regions (e.g., NGC\,5953, NGC\,7742).

\begin{figure}
\begin{center}
   \includegraphics[angle=0, width=0.99\linewidth]{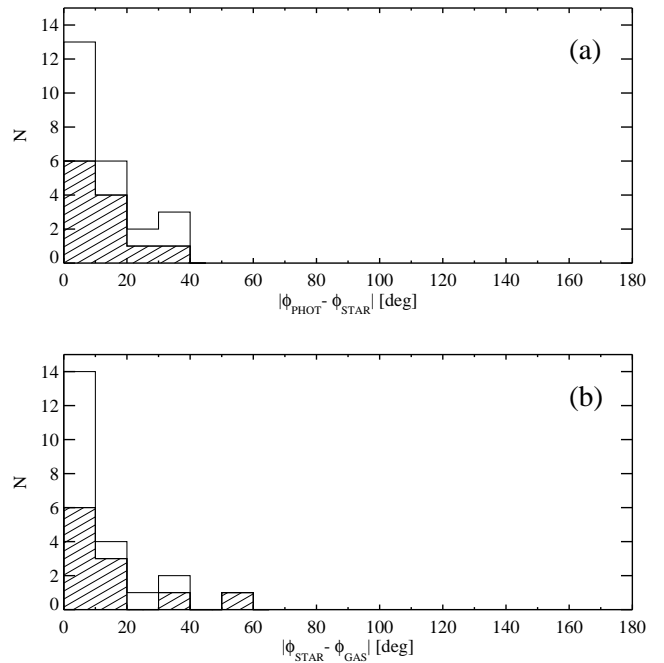}
\end{center}
\caption{Distribution of misalignments in our sample of galaxies. (a)
misalignment between the photometric and stellar kinematic major axes. (b)
misalignment between the stellar and gaseous kinematical major axes. In both 
panels the solid line represents the distribution for the whole sample, while 
the shaded area shows the distribution for field galaxies only.}
\label{fig:misalig}
\end{figure}

\subsection{Misalignments}
\label{subsec:misalignments}
The measurement of misalignments between structural components in early-type 
galaxies has often been used to assess the frequency of triaxiality, by 
measuring differences between the photometric and kinematic major axes (e.g., 
\citealt{binney85, fiz91}), or to determine the occurrence of accretion events, 
by measuring the misalignment between the kinematics of the stellar and gaseous
component (e.g., \citealt{kf01}). Here, we measure both types of misalignments 
to quantify the presence of bars or triaxial bulges and the presence of 
decoupled gaseous components in our sample, and to investigate their dependence 
on environment.

We determined the photometric major axis of each galaxy by computing the 
weighted second moments of the intensity distribution using the large scale 
ground-based MDM images (see Sect.~\ref{sec:observations}). This generally 
corresponds to the position angle of the main disc. We used the MDM images in 
the F814W filter to minimise the effects of dust in the determination of the 
photometric major axis. The stellar and gaseous kinematic major axes were 
obtained by finding the position angle that minimised the difference between the 
\sauron\ velocity fields and a bi-antisymmetric representation of the velocity 
field (see Appendix C in \citealt{davor06}). Both methods are robust indicators 
that provide a good estimate of the {\it global} position angle of the maps 
used, and thus peculiarities on small scales are not visible in the final
distribution of misalignments. The values of the measured misalignments are 
listed in Table~\ref{tab:allgal}.\looseness-2

The study of triaxiality due to past merging events is difficult to assess 
in our sample of disc galaxies. Spiral galaxies are known to suffer disc 
instabilities that could eventually lead to the formation of bars (i.e.,
internal triaxial bodies). With time, it is likely that the evolution of the bar
will erase any signature of triaxiality from any previous merging event. Despite
the progress in this field, the relative importance of internal versus external 
processes and their role on the formation of spiral galaxies is not well 
established yet (see \citealt{kormendy04} for a review on the subject).

In Figure~\ref{fig:misalig}(a), we quantify the degree of misalignment between 
the stellar surface brightness distribution with respect to the main stellar 
rotation axis for each galaxy. The histogram displays a gradually increasing 
number of galaxies towards small misalignments. As discussed in 
\S~\ref{subsec:stars}, bars are, at least in part, responsible for the spread 
towards intermediate misalignments. However, in the figure, barred galaxies 
alone do not account for the observed excess of galaxies at intermediate values. 
The presence of dust in many of the galaxies in our sample plays an important 
role in the observed distribution. No significant difference is found between 
field and cluster members. A KS-test indicates that there is a 78\% probability 
that the distributions of the two classes are identical.

Figure~\ref{fig:misalig}(b) presents a histogram with the observed misalignments 
between the kinematics of the stars and ionised gas. Similarly to 
panel~\ref{fig:misalig}(a) there is an excess of galaxies with small 
misalignments. The spread towards intermediate misalignments seems to be due to 
non-circular motions in the ionised-gas velocity fields (e.g., NGC\,5448, 
\citealt{fathi05}), and to a lesser extent by dust extinction. From these 
figures, the misalignment distribution appears to be consistent for field and 
cluster galaxies (KS-test, $p$=73\%). The only two galaxies without ionised-gas 
detection, NGC\,4425 and NGC\,6501, belong to cluster and field environments, 
respectively.\looseness-2

These results contrast with our findings for early-type galaxies (Paper V), as 
there are no cases displaying misalignments above 60$\degr$ in our sample. The 
absence of counter-rotating gas and stars or of gas minor-axis rotation in our 
sample is consistent with the finding of previous surveys (e.g., \citealt{kf01}) 
that such phenomena are exceedingly rare in spiral galaxies. A likely 
explanation for the observed distribution is that in the event of an accretion 
of a small satellite by a spiral galaxy, the axisymmetric geometry of the 
system will cause the accreted material to quickly settle on the main galactic 
plane, where pre-existing gas would shock with counter-rotating accreted gas, 
dissipating its angular momentum or inducing star formation (see also 
\citealt{rix95,corsini98,hjbbm00,kf01}, and references therein).

\section{Star formation in Sa bulges}
\label{sec:sfr}
Determining the rate at which stars are born in galaxies is key to understanding 
how galaxies form and evolve. Since the first estimates of star-formation rates 
using colors (e.g., \citealt{tinsley68}), many different diagnostics have been 
proposed to measure the amount of stellar mass produced in a given time (i.e., 
ultraviolet continuum, recombination lines, forbidden lines, far-infrared 
continuum). See \citet{kennicutt_rev} for an extensive review on this subject.
\looseness-2

In this section we use one of the most widely applied diagnostics to determine
the importance of the star formation rate (SFR) in our sample of galaxies: the
SFR$-$\halpha\ luminosity relation. We use the prescription of
\citet{kennicutt_rev}:

\begin{equation}
\rm{SFR}(\rm M_{\odot}~yr^{-1}) = 
{\frac{L(\halpha)}{7.9\times\ 10^{42}~erg~s^{-1}}}~~~.
\label{eq:sfr}
\end{equation}

\noindent The wavelength range covered by \sauron\ does not allow for a direct 
measurement of the \halpha\ luminosity, but \hbeta\ fluxes instead. We 
used the measured Hubble flow\footnote{$H_0$=71 \kms\ Mpc$^{-1}$ 
\citep{bennett03}.}, listed in Table~\ref{tab:allgal}, to convert our observed
\hbeta\ fluxes into luminosities. Given that many galaxies in our sample display 
significant amounts of dust, we corrected those \hbeta\ luminosities for dust 
attenuation. For the galactic extinction we used the correction factors in 
$V$-band of \citet{schlegel98} as given by the NED database. We also considered 
the expected dust extinction of each galaxy due to its inclination. For that we 
used the values listed in the Hyperleda catalogue, which were determined using 
the prescription of \citet{bottinelli95}. The values, originally in $B$-band, 
were converted to $V$-band assuming the attenuation law of \citet{cardelli89}. 
After these corrections, we converted the extinction corrected \hbeta\ 
luminosities to \halpha\ luminosities using the theoretical conversion factor 
2.86 \citep{osterbrock}. The SFR was then determined using the equation above. 
Finally, we derived the masses of the ionised gas in each galaxy following 
\citet{kim89} and assuming a Case B recombination at T=10,000K with an electron 
density of $n=100$ cm$^{-3}$ \citep{osterbrock}. Table~\ref{tab:allgal} lists 
the values of the observed \hbeta\ fluxes, correction factors for dust 
extinction, computed extinction corrected \halpha\ luminosities, star formation 
rates and ionised-gas masses within the \sauron\ FoV.

\begin{figure}
\begin{center}
   \includegraphics[angle=0, width=0.99\linewidth]{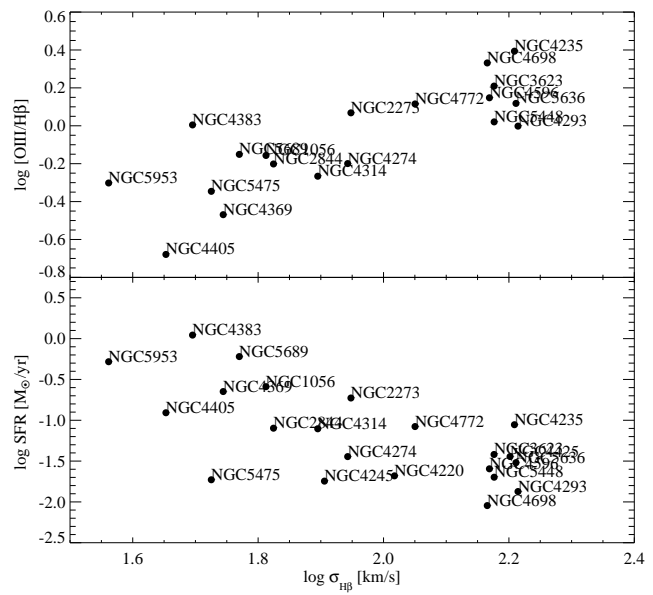}
\end{center}
\caption{\oiii/\hbeta\ ratio and star-formation rate versus \hbeta\ velocity 
dispersion. Galaxies displaying low velocity dispersions also show the lowest 
values of \oiii/\hbeta, suggesting a link between the level of star formation 
and the dynamical state of the gas. This is confirmed in the bottom panel where 
the higher star formation rates are found in galaxies with lowest \hbeta\ 
velocity dispersion.}
\label{fig:sfr}
\end{figure}

Following the investigation of the relation between star formation and low
ionised-gas velocity dispersions in sigma-drops in \S\ref{subsec:stars}, we
looked at whether a similar empirical relation holds in general in our sample.
In Figure \ref{fig:sfr} we plot \oiii/\hbeta\ ratio and SFR versus the \hbeta\ 
velocity dispersion for each galaxy. The \oiii/\hbeta\ ratio is the average over 
the \sauron\ FoV, while the SFR is the total over the same area. The \hbeta\ 
velocity dispersion is the luminosity weighted average of the individual 
measurements over the full \sauron\ map for each galaxy. We find that 
ionised-gas velocity dispersion values are low ($\sim$50 \kms) in regions where 
the \oiii/\hbeta\ ratio is also low ($\le$1, see Fig.\ref{fig:sfr}, top panel). 
Galaxies displaying the largest SFR are also those showing the lowest \hbeta\ 
velocity dispersions (see Fig.\ref{fig:sfr}, bottom panel). The obvious 
interpretation is that we are seeing stars being formed from cold (i.e., low 
velocity dispersion) gas. A similar behaviour is found in \sauron\ observations 
of the grand-design spiral galaxy M\,100 \citep{allard05} and \gmos\ IFU
observations of NGC\,1097 \citep{fathi06}. The relation, although still
present, becomes much weaker if the \oiii\ velocity dispersion is used instead,
displaying the limitations of the \oiii\ line as an indicator of star formation.

The star formation rates found here range over several orders of magnitude
and appear to be in good agreement with those for normal disc galaxies
\citep{kennicutt_rev}. Some of the most extreme cases of star formation in our
sample appear in the form of circumnuclear star-forming rings. In total we have
detected six (NGC\,2844, NGC\,4245, NGC\,4274, NGC\,4314, NGC\,5953, NGC\,7742)
in our sample of 24 galaxies, which is consistent with the most recent estimates
from \halpha\ studies of larger samples (21$\pm$5\%, \citealt{knapen05}). The
formation of star-forming rings is usually associated with the interplay between
bar-driven inflow and bar resonances \citep{schwarz84,byrd94,pst95}. In our
sample there are three galaxies, displaying inner rings, that are indeed
classified as barred (NGC\,4245, NGC\,4274, NGC\,4314). Star-forming rings in
general, however, do not need to be accompanied by the visible presence of a
bar. As gas in galaxies is very sensitive to non-axisymmetric structures, a
simple oval (i.e. weak bar) suffices to force the re-distribution of the gas
(i.e., \citealt{bc96}). NGC\,7742 (see also Paper II) could be an example of
this scenario, where there is no clear evidence of a bar. This hypothesis is
further supported by the presence of a central decoupled component in the inner
parts of the galaxy which, as discussed in Section~\ref{sec:results}, is common
in barred galaxies. However, given the striking resemblance of NGC\,7742 to
Hoag's Object \citep{schweizer87}, alternative formation scenarios may include 
an interaction event with a neighbour galaxy \citep{toomre77,hc01}. In this 
context, the star-forming ring in NGC\,2844 may be the consequence of an 
interaction with the nearby galaxies (e.g., NGC\,2852, NGC\,2853, see 
Appendix~\ref{sec:galaxies_notes}). An even more evident case of the interacting 
scenario is the close pair formed by NGC\,5953 and NGC\,5954. In this case, the 
star-forming ring in NGC\,5953 is co-spatial with the kinematically decoupled 
component seen in the stellar and gas kinematics (\S\ref{subsec:stars}, 
\S\ref{subsec:gas}), thus indicating that we may be witnessing the formation 
of a decoupled component as a result of an ongoing interaction.

\section{Concluding remarks}
\label{sec:conclusions}
The maps presented in this paper constitute a systematic study of the stellar
and ionised-gas morphology and kinematics of a representative sample of 24 Sa
galaxies in the nearby universe. Our analysis reveals stellar kinematically
decoupled components in 12 of the galaxies in our sample. Many of them show
additional velocity dispersion drops, which are often associated with the
presence of young stars. The ionised gas is almost ubiquitous in our sample and 
is, in general, is accompanied by the presence of dust. The kinematics of the 
gas is consistent with circular rotation in a disc co-rotating with respect to 
the stars. We assessed the importance of triaxiality and interactions 
by studying the distribution of misalignments between the photometric and 
stellar kinematics major axes, and also between the stellar and gas kinematic 
major axes. We concluded that, even though dust contamination plays an important 
role, bars are the main source for the observed non-axisymmetry in our sample. 
There is also no evidence for major accretion events in our sample, except in 
NGC\,4698, and NGC\,5953. Derived star formation rates appear to be in good 
agreement with previous work in the literature. In general, star formation is 
intense in many galaxies and displays varied morphologies; the most striking 
cases being six galaxies showing circumnuclear star-forming rings. We related 
the origin of these rings to gas redistribution by bars or ovals (e.g., 
NGC\,4314, NGC\,7742), or to interaction events (e.g., NGC\,5953).

The maps and analysis shown here comprise only part of the information that can
be extracted from the \sauron\ data. A more detailed analysis and interpretation
of some of the issues raised in this paper (i.e. including an in-depth study of
the stellar populations of these objects) will be the subject of a forthcoming
paper in this series.\looseness-2

\section*{Acknowledgments}
We would like to thank John Beckman and Almudena Zurita for very useful
discussions during the course of this paper, and also Katia Ganda and Glenn van
de Ven for their comments and a careful reading of the manuscript. We are grateful
to the referee, Daniel Thomas, for useful suggestions that have helped to 
improve the presentation of the main results of the paper. We thank the Isaac 
Newton Group staff, in particular Rene Rutten, Tom Gregory and Chris Benn, for 
enthusiastic and competent support on La Palma. JFB acknowledges support from 
the Euro3D Research Training Network, funded by the EC under contract 
HPRN-CT-2002-00305. MC acknowledges support from a VENI grant 639.041.203 
awarded by the Netherlands Organization for Scientific Research (NWO). KF
acknowledges financial support from the Wenner-Gren Foundations. The \sauron\
project is made possible through grants 614.13.003, 781.74.203, 614.000.301 and
614.031.015 from NWO and financial contributions from the Institut National des
Sciences del'Univers, the Universit\'e Claude Bernard Lyon~I, the Universities
of Durham, Leiden, and Oxford, the British Council, PPARC grant `Extragalactic
Astronomy \& Cosmology at Durham 1998--2002', and the Netherlands Research
School for Astronomy NOVA. RLD is grateful for the award of a PPARC Senior
Fellowship (PPA/Y/S/1999/00854) and postdoctoral support through PPARC grant
PPA/G/S/2000/00729. This project made use of the HyperLeda and NED databases.
The Digitized Sky Surveys were produced at the Space Telescope Science Institute
under U.S. Government grant NAG W-2166. The images of these surveys are based on
photographic data obtained using the Oschin Schmidt Telescope on Palomar
Mountain and the UK Schmidt Telescope.

%

\begin{thebibliography}{}

\bibitem[\protect\citeauthoryear{{Afanasiev} \& {Sil'chenko}}{{Afanasiev} \& {Sil'chenko}}{2005}]{as05}
{Afanasiev} V.~L.,  {Sil'chenko} O.~K.,  2005, \aap, 429, 825

\bibitem[\protect\citeauthoryear{{Allard}, {Peletier} \& {Knapen}}{{Allard}  et~al.}{2005}]{allard05}
{Allard} E.~L.,  {Peletier} R.~F.,    {Knapen} J.~H.,  2005, \apjl, 633, L25

\bibitem[\protect\citeauthoryear{{Asif}, {Mundell} \& {Pedlar}}{{Asif} et~al.}{2005}]{amp05}
{Asif} M.~W.,  {Mundell} C.~G.,    {Pedlar} A.,  2005, \mnras, 359, 408

\bibitem[\protect\citeauthoryear{{Bacon} et~al.}{{Bacon} et~al.}{2001}]{bacon01}
{Bacon} R., {et~al.,} 2001, \mnras, 326, 23 [Paper I]

\bibitem[\protect\citeauthoryear{{Balcells} \& {Peletier}}{{Balcells} \& {Peletier}}{1994}]{bp94}
{Balcells} M.,  {Peletier} R.~F.,  1994, \aj, 107, 135

\bibitem[\protect\citeauthoryear{{Benedict}, {Higdon}, {Tollestrup}, {Hahn} \& {Harvey}}{{Benedict} et~al.}{1992}]{benedict92}
{Benedict} G.~F.,  {Higdon} J.~L.,  {Tollestrup} E.~V.,  {Hahn} J.~M., {Harvey} P.~M.,  1992, \aj, 103, 757

\bibitem[\protect\citeauthoryear{{Benedict}, {Smith} \& {Kenney}}{{Benedict} et~al.}{1996}]{bsk96}
{Benedict} G.~F.,  {Smith} B.~J.,    {Kenney} J.~D.~P.,  1996, \aj, 112, 1318

\bibitem[\protect\citeauthoryear{{Bennett} et al.}{{Bennett} et al.}{2003}]{bennett03}
{Bennett} C.~L.,  {et al.,} 2003, \apjs, 148, 97

\bibitem[\protect\citeauthoryear{{Bernardi}, {Alonso}, {da Costa}, {Willmer}, {Wegner}, {Pellegrini}, {Rit{\' e}} \& {Maia}}{{Bernardi} et~al.}{2002}]{bernardi02}
{Bernardi} M.,  {Alonso} M.~V.,  {da Costa} L.~N.,  {Willmer} C.~N.~A.,
  {Wegner} G.,  {Pellegrini} P.~S.,  {Rit{\' e}} C.,    {Maia} M.~A.~G.,  2002,
  \aj, 123, 2990

\bibitem[\protect\citeauthoryear{{Bertola}, {Corsini}, {Vega Beltr{\' a}n}, {Pizzella}, {Sarzi}, {Cappellari} \& {Funes}}{{Bertola} et~al.}{1999}]{bertola99}
{Bertola} F.,  {Corsini} E.~M.,  {Vega Beltr{\' a}n} J.~C.,  {Pizzella} A.,
  {Sarzi} M.,  {Cappellari} M.,    {Funes} J.~G.,  1999, \apjl, 519, L127

\bibitem[\protect\citeauthoryear{{Binney}}{{Binney}}{1980}]{binney80}
{Binney} J.,  1980, \mnras, 190, 873

\bibitem[\protect\citeauthoryear{{Binney}}{{Binney}}{1985}]{binney85}
{Binney} J.,  1985, \mnras, 212, 767

\bibitem[\protect\citeauthoryear{{Bottema}}{{Bottema}}{1989}]{bottema89}
{Bottema} R.,  1989, \aap, 221, 236

\bibitem[\protect\citeauthoryear{{Bottema}}{{Bottema}}{1993}]{bottema93}
{Bottema} R.,  1993, \aap, 275, 16

\bibitem[\protect\citeauthoryear{{Bottinelli}, {Gouguenheim}, {Paturel} \& {Teerikorpi}}{{Bottinelli}      et~al.}{1995}]{bottinelli95}
{Bottinelli} L.,  {Gouguenheim} L.,  {Paturel} G.,    {Teerikorpi} P.,  1995, \aap, 296, 64

\bibitem[\protect\citeauthoryear{{Bureau} \& {Athanassoula}}{{Bureau} \& {Athanassoula}}{2005}]{ba05}
{Bureau} M.,  {Athanassoula} E.,  2005, \apj, 626, 159

\bibitem[\protect\citeauthoryear{{Buta} \& {Combes}}{{Buta} \& {Combes}}{1996}]{bc96}
{Buta} R.,  {Combes} F.,  1996, Fundamentals of Cosmic Physics, 17, 95

\bibitem[\protect\citeauthoryear{{Byrd}, {Rautiainen}, {Salo}, {Buta} \& {Crocher}}{{Byrd} et~al.}{1994}]{byrd94}
{Byrd} G.,  {Rautiainen} P.,  {Salo} H.,  {Buta} R.,    {Crocher} D.~A.,  1994, \aj, 108, 476

\bibitem[\protect\citeauthoryear{{Cappellari} \& {Copin}}{{Cappellari} \& {Copin}}{2003}]{michele03}
{Cappellari} M.,  {Copin} Y.,  2003, \mnras, 342, 345

\bibitem[\protect\citeauthoryear{{Cappellari} \& {Emsellem}}{{Cappellari} \& {Emsellem}}{2004}]{capem04}
{Cappellari} M.,  {Emsellem} E.,  2004, \pasp, 116, 138

\bibitem[\protect\citeauthoryear{{Cardelli}, {Clayton} \& {Mathis}}{{Cardelli} et~al.}{1989}]{cardelli89}
{Cardelli} J.~A.,  {Clayton} G.~C.,    {Mathis} J.~S.,  1989, \apj, 345, 245

\bibitem[\protect\citeauthoryear{{Chemin} et al.}{{Chemin} et al.}{2005}]{chemin05}
{Chemin} L.,  {et al.,} 2005, \aap, 436, 469

\bibitem[\protect\citeauthoryear{{Chemin} et al.}{{Chemin} et al.}{2006}]{chemin06}
{Chemin} L.,  {et al.,} 2006, \mnras, 366, 812

\bibitem[\protect\citeauthoryear{{Chromey}, {Elmegreen}, {Mandell} \& {McDermott}}{{Chromey} et~al.}{1998}]{cemd98}
{Chromey} F.~R.,  {Elmegreen} D.~M.,  {Mandell} A.,    {McDermott} J.,  1998, \aj, 115, 2331

\bibitem[\protect\citeauthoryear{{Chung} \& {Bureau}}{{Chung} \& {Bureau}}{2004}]{cb04}
{Chung} A.,  {Bureau} M.,  2004, \aj, 127, 3192

\bibitem[\protect\citeauthoryear{{Cid Fernandes} et al.}{{Cid Fernandes} et~al.}{2004}]{cid04}
{Cid Fernandes} R., et al.,  2004, \apj, 605, 105

\bibitem[\protect\citeauthoryear{{Ciotti} \& {Lanzoni}}{{Ciotti} \& {Lanzoni}}{1997}]{cl97}
{Ciotti} L.,  {Lanzoni} B.,  1997, \aap, 321, 724

\bibitem[\protect\citeauthoryear{{Corsini}, {Pizzella}, {Coccato} \& {Bertola}}{{Corsini} et~al.}{2003}]{corsini03}
{Corsini} E.~M.,  {Pizzella} A.,  {Coccato} L.,    {Bertola} F.,  2003, \aap, 408, 873

\bibitem[\protect\citeauthoryear{{Corsini}, {Pizzella}, {Funes}, {Vega Beltran} \& {Bertola}}{{Corsini} et~al.}{1998}]{corsini98}
{Corsini} E.~M.,  {Pizzella} A.,  {Funes} J.~G.,  {Vega Beltran} J.~C., {Bertola} F.,  1998, \aap, 337, 80

\bibitem[\protect\citeauthoryear{{de Vaucouleurs}}{{de Vaucouleurs}}{1948}]{devauc48}
{de Vaucouleurs} G.,  1948, Annales d'Astrophysique, 11, 247

\bibitem[\protect\citeauthoryear{{de Zeeuw} et al.}{{de Zeeuw} et~al.}{2002}]{tim02}
{de Zeeuw} P.~T., {et al.,} 2002, \mnras, 329, 513 [Paper II]

\bibitem[\protect\citeauthoryear{{di Nella}, {Garcia}, {Garnier} \& {Paturel}}{{di Nella} et~al.}{1995}]{dinella95}
{di Nella} H.,  {Garcia} A.~M.,  {Garnier} R.,    {Paturel} G.,  1995, \aaps, 113, 151

\bibitem[\protect\citeauthoryear{{Emsellem} et al.}{{Emsellem} et~al.}{2004}]{emsellem04}
{Emsellem} E., {et al.,} 2004, \mnras, 352, 721 [Paper III]

\bibitem[\protect\citeauthoryear{{Emsellem} et al.}{{Emsellem} et~al.}{2001}]{emsellem01}
{Emsellem} E., {Greusard} D., {Combes} F., {Friedli} D., {Leon} S., {P{\'e}contal} E., {Wozniak} H.,  2001, \aap, 368, 52

\bibitem[\protect\citeauthoryear{{Erwin}}{{Erwin}}{2004}]{erwin04}
{Erwin} P.,  2004, \aap, 415, 941

\bibitem[\protect\citeauthoryear{{Erwin} \& {Sparke}}{{Erwin} \& {Sparke}}{2003}]{es03}
{Erwin} P.,  {Sparke} L.~S.,  2003, \apjs, 146, 299

\bibitem[\protect\citeauthoryear{{Eskridge} et al.}{{Eskridge} et~al.}{2002}]{eskridge02}
{Eskridge} P.~B.,  et al., 2002, \apjs, 143, 73

\bibitem[\protect\citeauthoryear{{Falc{\'o}n-Barroso}, {Balcells}, {Peletier} \& {Vazdekis}}{{Falc{\'o}n-Barroso} et~al.}{2003}]{fbpv03}
{Falc{\'o}n-Barroso} J.,  {Balcells} M.,  {Peletier} R.~F.,    {Vazdekis} A., 2003, \aap, 405, 455

\bibitem[\protect\citeauthoryear{{Fathi}, {van de Ven}, {Peletier}, {Emsellem}, {Falc\'on-Barroso}, {Cappellari} \& {de Zeeuw}}{{Fathi} et~al.}{2005}]{fathi05}
{Fathi} K.,  {van de Ven} G.,  {Peletier} R.,  {Emsellem} E., 
{Falc\'on-Barroso} J.,  {Cappellari} M.,    {de Zeeuw} P.,  2005, \mnras, 364,
773 

\bibitem[\protect\citeauthoryear{{Fathi},{Storchi-Bergmann},{Riffel},{Winge},
{Axon},{Robinson},{Capetti},{Marconi}}{{Fathi} et~al.}{2006}]{fathi06}
{Fathi} K., {Storchi-Bergmann} T., {Riffel} R.~A., {Winge} C., {Axon} D.~J.,
{Robinson} A., {Capetti} A., {Marconi} A., 2006, \apjl, in press

\bibitem[\protect\citeauthoryear{{Ferruit}, {Binette}, {Sutherland} \& {Pecontal}}{{Ferruit} et~al.}{1997}]{fbsp97}
{Ferruit} P.,  {Binette} L.,  {Sutherland} R.~S.,    {Pecontal} E.,  1997, \aap, 322, 73

\bibitem[\protect\citeauthoryear{{Ferruit}, {Wilson} \& {Mulchaey}}{{Ferruit} et~al.}{2000}]{fwm00}
{Ferruit} P.,  {Wilson} A.~S.,    {Mulchaey} J.,  2000, \apjs, 128, 139

\bibitem[\protect\citeauthoryear{{Fisher}}{{Fisher}}{1997}]{fisher97}
{Fisher} D.,  1997, \aj, 113, 950

\bibitem[\protect\citeauthoryear{{Franx}, {Illingworth} \& {de Zeeuw}}{{Franx} et~al.}{1991}]{fiz91}
{Franx} M.,  {Illingworth} G.,    {de Zeeuw} T.,  1991, \apj, 383, 112

\bibitem[\protect\citeauthoryear{{Friedli} \& {Benz}}{{Friedli} \& {Benz}}{1995}]{fb95}
{Friedli} D.,  {Benz} W.,  1995, \aap, 301, 649

\bibitem[\protect\citeauthoryear{{Ganda}, {Falc\'on-Barroso}, {Peletier},
  {Cappellari}, {Emsellem}, {McDermid}, {de Zeeuw} \& {Carollo}}{{Ganda}
  et~al.}{2005}]{ganda05}
{Ganda} K.,  {Falc\'on-Barroso} J.,  {Peletier} R.~F.,  {Cappellari} M.,
  {Emsellem} E.,  {McDermid} R.~M.,  {de Zeeuw} P.~T.,  {Carollo} C.~M.,
  2006, \mnras, in press (astro-ph/0512304)

\bibitem[\protect\citeauthoryear{{Gerhard}}{{Gerhard}}{1993}]{gerhard93}
{Gerhard} O.~E.,  1993, \mnras, 265, 213

\bibitem[\protect\citeauthoryear{{Gerssen}, {Kuijken} \& {Merrifield}}{{Gerssen} et~al.}{1999}]{gkm99}
{Gerssen} J.,  {Kuijken} K.,    {Merrifield} M.~R.,  1999, \mnras, 306, 926

\bibitem[\protect\citeauthoryear{{Giuricin}, {Marinoni}, {Ceriani} \& {Pisani}}{{Giuricin} et~al.}{2000}]{gmcp00}
{Giuricin} G.,  {Marinoni} C.,  {Ceriani} L.,    {Pisani} A.,  2000, \apj, 543, 178\looseness-2

\bibitem[\protect\citeauthoryear{{Gonzalez Delgado} \& {Perez}}{{Gonzalez Delgado} \& {Perez}}{1996}]{gp96}
{Gonzalez Delgado} R.~M.,  {Perez} E.,  1996, \mnras, 281, 781

\bibitem[\protect\citeauthoryear{{Gonzalez Delgado}, {Perez}, {Tadhunter}, 
{Vilchez} \& {Rodriguez-Espinosa}}{{Gonzalez Delgado} et~al.}{1997}]{gptvr97}
{Gonzalez Delgado} R.~M.,  {Perez} E.,  {Tadhunter} C.,  {Vilchez} J.~M.,
  {Rodriguez-Espinosa} J.~M.,  1997, \apjs, 108, 155

\bibitem[\protect\citeauthoryear{{Goudfrooij} \& {Emsellem}}{{Goudfrooij} \& {Emsellem}}{1996}]{ge96}
{Goudfrooij} P.,  {Emsellem} E.,  1996, \aap, 306, L45

\bibitem[\protect\citeauthoryear{{Hameed} \& {Devereux}}{{Hameed} \& {Devereux}}{2005}]{hameed05}
{Hameed} S.,  {Devereux} N.,  2005, \aj, 129, 2597

\bibitem[\protect\citeauthoryear{{Haynes}, {Jore}, {Barrett}, {Broeils} \& 
{Murray}}{{Haynes} et~al.}{2000}]{hjbbm00}
{Haynes} M.~P.,  {Jore} K.~P.,  {Barrett} E.~A.,  {Broeils} A.~H.,    {Murray}
  B.~M.,  2000, \aj, 120, 703

\bibitem[\protect\citeauthoryear{{Heller} \& {Shlosman}}{{Heller} \& {Shlosman}}{1994}]{hs94}
{Heller} C.~H.,  {Shlosman} I.,  1994, \apj, 424, 84

\bibitem[\protect\citeauthoryear{{Hern{\' a}ndez-Toledo}, {Fuentes-Carrera},
  {Rosado}, {Cruz-Gonz{\' a}lez}, {Franco-Balderas} \&
  {Dultzin-Hacyan}}{{Hern{\' a}ndez-Toledo} et~al.}{2003}]{hernandez03}
{Hern{\' a}ndez-Toledo} H.~M.,  {Fuentes-Carrera} I.,  {Rosado} M.,
  {Cruz-Gonz{\' a}lez} I.,  {Franco-Balderas} A.,    {Dultzin-Hacyan} D.,
  2003, \aap, 412, 669

\bibitem[\protect\citeauthoryear{{Ho}, {Filippenko} \& {Sargent}}{{Ho} et~al.}{1995}]{ho95}
{Ho} L.~C.,  {Filippenko} A.~V.,    {Sargent} W.~L.,  1995, \apjs, 98, 477

\bibitem[\protect\citeauthoryear{{Ho}, {Filippenko} \& {Sargent}}{{Ho} et~al.}{1997a}]{ho97}
{Ho} L.~C.,  {Filippenko} A.~V.,    {Sargent} W.~L.~W.,  1997a, \apjs, 112, 315

\bibitem[\protect\citeauthoryear{{Ho}, {Filippenko}, {Sargent} \& {Peng}}{{Ho} et~al.}{1997b}]{ho97b}
{Ho} L.~C.,  {Filippenko} A.~V.,  {Sargent} W.~L.~W.,    {Peng} C.~Y.,  1997b, \apjs, 112, 391

\bibitem[\protect\citeauthoryear{{Horellou} \& {Combes}}{{Horellou} \& {Combes}}{2001}]{hc01}
{Horellou} C.,  {Combes} F.,  2001, \apss, 276, 1141

\bibitem[\protect\citeauthoryear{{Jenkins}}{{Jenkins}}{1984}]{jenkins84}
{Jenkins} C.~R.,  1984, \apj, 277, 501

\bibitem[\protect\citeauthoryear{{Jim{\' e}nez-Benito}, {D{\'{\i}}az},
  {Terlevich} \& {Terlevich}}{{Jim{\' e}nez-Benito} et~al.}{2000}]{jdtt00}
{Jim{\' e}nez-Benito} L.,  {D{\'{\i}}az} A.~I.,  {Terlevich} R.,    {Terlevich}
  E.,  2000, \mnras, 317, 907

\bibitem[\protect\citeauthoryear{{Kannappan} \& {Fabricant}}{{Kannappan} \& {Fabricant}}{2001}]{kf01}
{Kannappan} S.~J.,  {Fabricant} D.~G.,  2001, \aj, 121, 140

\bibitem[\protect\citeauthoryear{{Kennicutt}}{{Kennicutt}}{1998}]{kennicutt_rev}
{Kennicutt} R.~C.,  1998, \araa, 36, 189

\bibitem[\protect\citeauthoryear{{Kim}}{{Kim}}{1989}]{kim89}
{Kim} D.-W.,  1989, \apj, 346, 653

\bibitem[\protect\citeauthoryear{{Knapen}}{{Knapen}}{2005}]{knapen05}
{Knapen} J.~H.,  2005, \aap, 429, 141

\bibitem[\protect\citeauthoryear{{Koopmann} \& {Kenney}}{{Koopmann} \& {Kenney}}{2004}]{kk04}
{Koopmann} R.~A.,  {Kenney} J.~D.~P.,  2004, \apj, 613, 866

\bibitem[\protect\citeauthoryear{{Kormendy}}{{Kormendy}}{1993}]{kormendy93}
{Kormendy} J.,  1993, in IAU Symp. 153: Galactic Bulges {Kinematics of
  extragalactic bulges: evidence that some bulges are really disks}. p.~209

\bibitem[\protect\citeauthoryear{Kormendy \& Kennicutt}{2004}]{kormendy04} 
Kormendy J., Kennicutt R.~C., 2004, ARA\&A, 42, 603 

\bibitem[\protect\citeauthoryear{{Krajnovi\' c}, {Cappellari}, {de Zeeuw} \& 
{Copin}}{{Krajnovi\' c} et~al.}{2006}]{davor06}
{Krajnovi\' c} D.,  {Cappellari} M.,  {de Zeeuw} P.,    {Copin} Y.,  2006, 
\mnras, 366, 787

\bibitem[\protect\citeauthoryear{{Kuijken} \& {Merrifield}}{{Kuijken} \& {Merrifield}}{1995}]{kuijken95}
{Kuijken} K.,  {Merrifield} M.~R.,  1995, \apjl, 443, L13

\bibitem[\protect\citeauthoryear{{Kuntschner} et al.}{{Kuntschner} et~al.}
{2006}]{kuntschner06}
{Kuntschner} H., et al., 2006, \mnras, in press [Paper VI] (astro-ph/0602192)

\bibitem[\protect\citeauthoryear{{L{\" u}tticke}, {Dettmar} \& {Pohlen}}{{L{\"u}tticke} et~al.}{2000}]{ldp00}
{L{\" u}tticke} R.,  {Dettmar} R.-J.,    {Pohlen} M.,  2000, \aaps, 145, 405

\bibitem[\protect\citeauthoryear{{Landsman}}{{Landsman}}{1993}]{landsman93}
{Landsman} W.~B.,  1993, in ASP Conf. Ser. 52: Astronomical Data Analysis
  Software and Systems II {The IDL Astronomy User's Library}. p.~246

\bibitem[\protect\citeauthoryear{{Laurikainen} \& {Salo}}{{Laurikainen} \& {Salo}}{2002}]{ls02}
{Laurikainen} E.,  {Salo} H.,  2002, \mnras, 337, 1118

\bibitem[\protect\citeauthoryear{{M{\'a}rquez}, {Masegosa}, {Durret},
  {Gonz{\'a}lez Delgado}, {Moles}, {Maza}, {P{\'e}rez} \& {Roth}}{{M{\'a}rquez}
  et~al.}{2003}]{marquez03}
{M{\'a}rquez} I.,  {Masegosa} J.,  {Durret} F.,  {Gonz{\'a}lez Delgado} R.~M.,
  {Moles} M.,  {Maza} J.,  {P{\'e}rez} E.,    {Roth} M.,  2003, \aap, 409, 459


\bibitem[\protect\citeauthoryear{{Osterbrock}}{{Osterbrock}}{1989}]{osterbrock}
{Osterbrock} D.~E.,  1989, {Astrophysics of gaseous nebulae and active galactic 
nuclei}. University Science Books, 1989, p. 422


\bibitem[\protect\citeauthoryear{{P{\' e}rez-Ram{\'{\i}}rez}, {Knapen},
  {Peletier}, {Laine}, {Doyon} \& {Nadeau}}{{P{\' e}rez-Ram{\'{\i}}rez}
  et~al.}{2000}]{pkpldn00}
{P{\' e}rez-Ram{\'{\i}}rez} D.,  {Knapen} J.~H.,  {Peletier} R.~F.,  {Laine}
  S.,  {Doyon} R.,    {Nadeau} D.,  2000, \mnras, 317, 234

\bibitem[\protect\citeauthoryear{{Peletier}, {Knapen}, {Shlosman},
  {P{\'e}rez-Ram{\'{\i}}rez}, {Nadeau}, {Doyon}, {Rodriguez Espinosa} \&
  {P{\'e}rez Garc{\'{\i}}a}}{{Peletier} et~al.}{1999}]{peletier99}
{Peletier} R.~F.,  {Knapen} J.~H.,  {Shlosman} I.,  {P{\'e}rez-Ram{\'{\i}}rez}
  D.,  {Nadeau} D.,  {Doyon} R.,  {Rodriguez Espinosa} J.~M.,    {P{\'e}rez
  Garc{\'{\i}}a} A.~M.,  1999, \apjs, 125, 363

\bibitem[\protect\citeauthoryear{{Piner}, {Stone} \& {Teuben}}{{Piner} et~al.}{1995}]{pst95}
{Piner} B.~G.,  {Stone} J.~M.,    {Teuben} P.~J.,  1995, \apj, 449, 508

\bibitem[\protect\citeauthoryear{{Pizzella}, {Corsini}, {Morelli}, {Sarzi},
  {Scarlata}, {Stiavelli} \& {Bertola}}{{Pizzella} et~al.}{2002}]{pizzella02}
{Pizzella} A.,  {Corsini} E.~M.,  {Morelli} L.,  {Sarzi} M.,  {Scarlata} C.,
  {Stiavelli} M.,    {Bertola} F.,  2002, \apj, 573, 131

\bibitem[\protect\citeauthoryear{{Plana}, {Boulesteix}, {Amram}, {Carignan} \&
  {Mendes de Oliveira}}{{Plana} et~al.}{1998}]{plana98}
{Plana} H.,  {Boulesteix} J.,  {Amram} P.,  {Carignan} C.,    {Mendes de
  Oliveira} C.,  1998, \aaps, 128, 75

\bibitem[\protect\citeauthoryear{{Press}, {Teukolsky}, {Vetterling} \&
  {Flannery}}{{Press} et~al.}{1992}]{press92}
{Press} W.~H.,  {Teukolsky} S.~A.,  {Vetterling} W.~T.,    {Flannery} B.~P.,
  1992, {Numerical recipes in FORTRAN. The art of scientific computing}.
Cambridge: University Press, c1992, 2nd ed.

\bibitem[\protect\citeauthoryear{{Prugniel} \& {Simien}}{{Prugniel} \& {Simien}}{1996}]{ps96}
{Prugniel} P.,  {Simien} F.,  1996, \aap, 309, 749

\bibitem[\protect\citeauthoryear{{Rampazzo}, {Reduzzi}, {Sulentic} \&
  {Madejsky}}{{Rampazzo} et~al.}{1995}]{rampazzo95}
{Rampazzo} R.,  {Reduzzi} L.,  {Sulentic} J.~W.,    {Madejsky} R.,  1995,
  \aaps, 110, 131

\bibitem[\protect\citeauthoryear{{Reshetnikov}}{{Reshetnikov}}{1993}]{res93}
{Reshetnikov} V.~P.,  1993, \aap, 280, 400

\bibitem[\protect\citeauthoryear{{Rix}, {Kennicutt}, {Braun} \&
  {Walterbos}}{{Rix} et~al.}{1995}]{rix95}
{Rix} H.-W.~R.,  {Kennicutt} R.~C.,  {Braun} R.,    {Walterbos} R.~A.~M.,
  1995, \apj, 438, 155

\bibitem[\protect\citeauthoryear{{Rubin}, {Waterman} \& {Kenney}}{{Rubin}
  et~al.}{1999}]{rwk99}
{Rubin} V.~C.,  {Waterman} A.~H.,    {Kenney} J.~D.~P.,  1999, \aj, 118, 236

\bibitem[\protect\citeauthoryear{{Sarzi}, {Corsini}, {Pizzella}, {Vega
  Beltr{\'a}n}, {Cappellari}, {Funes} \& {Bertola}}{{Sarzi}
  et~al.}{2000}]{sarzi00}
{Sarzi} M.,  {Corsini} E.~M.,  {Pizzella} A.,  {Vega Beltr{\'a}n} J.~C.,
  {Cappellari} M.,  {Funes} J.~G.,    {Bertola} F.,  2000, \aap, 360, 439

\bibitem[\protect\citeauthoryear{{Sarzi} et al.}{{Sarzi} et~al.}{2005}]{sarzi05}
{Sarzi} M.,  et al. 2005, \mnras, 366, 1151 [Paper V]

\bibitem[\protect\citeauthoryear{{Schlegel}, {Finkbeiner} \&
  {Davis}}{{Schlegel} et~al.}{1998}]{schlegel98}
{Schlegel} D.~J.,  {Finkbeiner} D.~P.,    {Davis} M.,  1998, \apj, 500, 525

\bibitem[\protect\citeauthoryear{{Schwarz}}{{Schwarz}}{1984}]{schwarz84}
{Schwarz} M.~P.,  1984, \mnras, 209, 93

\bibitem[\protect\citeauthoryear{{Schweizer}, {Ford}, {Jederzejewski} \&
  {Giovanelli}}{{Schweizer} et~al.}{1987}]{schweizer87}
{Schweizer} F.,  {Ford} W.~K.~J.,  {Jederzejewski} R.,    {Giovanelli} R.,
  1987, \apj, 320, 454

\bibitem[\protect\citeauthoryear{{Sersic}}{{Sersic}}{1968}]{sersic}
{Sersic} J.~L.,  1968, {Atlas de Galaxias Australes}. Cordoba, Argentina: 
Observatorio Astronomico.

\bibitem[\protect\citeauthoryear{{Shaw}, {Axon}, {Probst} \& {Gatley}}{{Shaw}
  et~al.}{1995}]{sapg95}
{Shaw} M.,  {Axon} D.,  {Probst} R.,    {Gatley} I.,  1995, \mnras, 274, 369

\bibitem[\protect\citeauthoryear{{Sofue}, {Tomita}, {Tutui}, {Honma} \&
  {Takeda}}{{Sofue} et~al.}{1998}]{sttht98}
{Sofue} Y.,  {Tomita} A.,  {Tutui} Y.,  {Honma} M.,    {Takeda} Y.,  1998,
  \pasj, 50, 427

\bibitem[\protect\citeauthoryear{{Terlevich}, {Diaz} \&
  {Terlevich}}{{Terlevich} et~al.}{1990}]{terlevich90}
{Terlevich} E.,  {Diaz} A.~I.,    {Terlevich} R.,  1990, \mnras, 242, 271

\bibitem[\protect\citeauthoryear{{Tinsley}}{{Tinsley}}{1968}]{tinsley68}
{Tinsley} B.~M.,  1968, \apj, 151, 547

\bibitem[\protect\citeauthoryear{{Toomre}}{{Toomre}}{1977}]{toomre77}
{Toomre} A.,  1977, p.~401

\bibitem[\protect\citeauthoryear{{Usui}, {Saito} \& {Tomita}}{{Usui}
  et~al.}{1998}]{ust98}
{Usui} T.,  {Saito} M.,    {Tomita} A.,  1998, \aj, 116, 2166

\bibitem[\protect\citeauthoryear{{van der Marel} \& {Franx}}{{van der Marel} \&
  {Franx}}{1993}]{vdm93}
{van der Marel} R.~P.,  {Franx} M.,  1993, \apj, 407, 525

\bibitem[\protect\citeauthoryear{{Vazdekis}}{{Vazdekis}}{1999}]{vazdekis99}
{Vazdekis} A.,  1999, \apj, 513, 224

\bibitem[\protect\citeauthoryear{{Veilleux}, {Kim}, {Sanders}, {Mazzarella} \&
  {Soifer}}{{Veilleux} et~al.}{1995}]{veilleux95}
{Veilleux} S.,  {Kim} D.-C.,  {Sanders} D.~B.,  {Mazzarella} J.~M.,    {Soifer}
  B.~T.,  1995, \apjs, 98, 171

\bibitem[\protect\citeauthoryear{{Wozniak}, {Combes}, {Emsellem} \&
  {Friedli}}{{Wozniak} et~al.}{2003}]{wozniak03}
{Wozniak} H.,  {Combes} F.,  {Emsellem} E.,    {Friedli} D.,  2003, \aap, 409,
  469

\bibitem[\protect\citeauthoryear{{Yoshida}, {Yamada}, {Kosugi}, {Taniguchi} \&
  {Mouri}}{{Yoshida} et~al.}{1993}]{yyktm93}
{Yoshida} M.,  {Yamada} T.,  {Kosugi} G.,  {Taniguchi} Y.,    {Mouri} H.,
  1993, \pasj, 45, 761

\end{thebibliography}
\bibliographystyle{mn2e} 



%
%
\renewcommand{\thefigure}{\arabic{figure}\alph{subfigure}}
\setcounter{figure}{5}
\setcounter{subfigure}{1}

\clearpage
\begin{figure*}
\begin{center}
  \includegraphics[angle=00,width=0.95\linewidth]{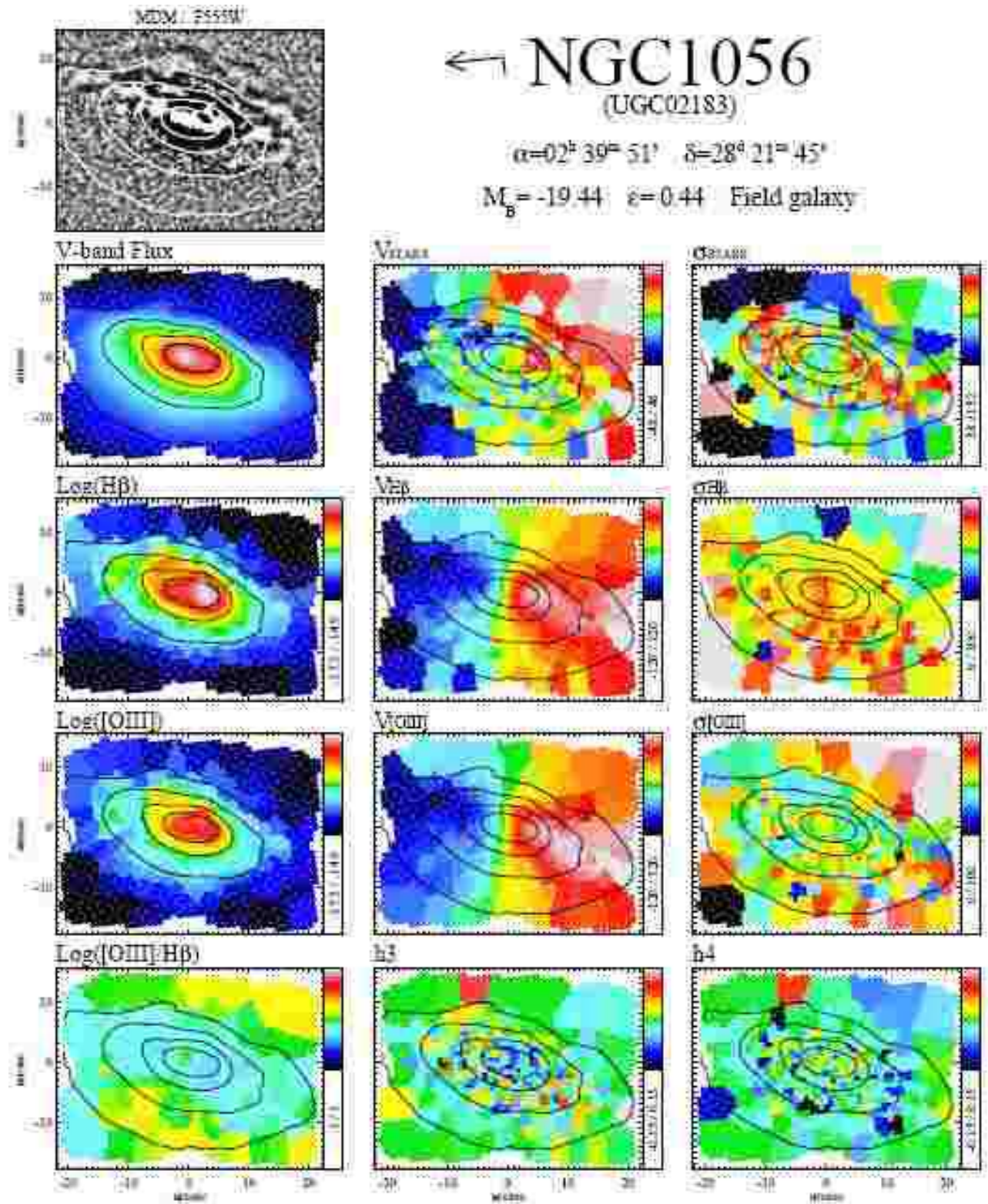}
\end{center}
\caption{Maps of the stellar and ionised-gas distribution and kinematics
of the 24 Sa galaxies in the \sauron\ representative sample. The \sauron\
spectra have been spatially binned to a minimum signal-to-noise of 60 by means
of the Voronoi 2D-binning algorithm of \citet{michele03}. The arrow and its 
associated dash at the top of each figure mark the North and East directions, 
respectively; the corresponding position angle of the vertical (upward) axis is 
provided in Table~\ref{tab:allgal}. {\it First row:} HST or ground-based 
unsharp-masked image. {\it Second row:} (i) reconstructed total intensity (in 
mag/arcsec$^2$ with an arbitrary zero point), (ii) stellar mean velocity $V$, 
and (iii) stellar velocity dispersion $\sigma$ (in \kms). {\it Third row:} (i)
\hbeta\ flux (in logarithmic scale), (ii) mean radial velocity, and (iii) 
velocity dispersion (in \kms). {\it Fourth row:} as {\it third row} for the 
\oiii\ line. {\it Fifth row:} (i) \oiii/\hbeta\ ratio map (in logarithmic 
scale), (ii) and (iii) Gauss--Hermite moments $h_3$ and $h_4$. The cuts levels 
are indicated in a box on the right hand side of each map. Those bins in the 
ionised-gas maps with $A/N$ below the thresholds stated in the text are colored 
in dark grey.}
\label{fig:N1056}
\end{figure*}

\addtocounter{figure}{-1}
\addtocounter{subfigure}{1}

\clearpage
\begin{figure*}
\begin{center}
  \includegraphics[angle=0, width=0.95\linewidth]{NGC2273_maps.ps2}
\end{center}
\caption{}
\label{fig:N2273}
\end{figure*}

\addtocounter{figure}{-1}
\addtocounter{subfigure}{1}

\clearpage
\begin{figure*}
\begin{center}
  \includegraphics[angle=0, width=0.95\linewidth]{NGC2844_maps.ps2}
\end{center}
\caption{}
\label{fig:N2844}
\end{figure*}

\addtocounter{figure}{-1}
\addtocounter{subfigure}{1}

\clearpage
\begin{figure*}
\begin{center}
  \includegraphics[angle=0, width=0.95\linewidth]{NGC3623_maps.ps2}
\end{center}
\caption{}
\label{fig:N3623}
\end{figure*}

\addtocounter{figure}{-1}
\addtocounter{subfigure}{1}

\clearpage
\begin{figure*}
\begin{center}
  \includegraphics[angle=0, width=0.95\linewidth]{NGC4220_maps.ps2}
\end{center}
\caption{}
\label{fig:N4420}
\end{figure*}

\addtocounter{figure}{-1}
\addtocounter{subfigure}{1}

\clearpage
\begin{figure*}
\begin{center}
  \includegraphics[angle=0, width=0.95\linewidth]{NGC4235_maps.ps2}
\end{center}
\caption{}
\label{fig:N4235}
\end{figure*}

\addtocounter{figure}{-1}
\addtocounter{subfigure}{1}

\clearpage
\begin{figure*}
\begin{center}
  \includegraphics[angle=0, width=0.95\linewidth]{NGC4245_maps.ps2}
\end{center}
\caption{}
\label{fig:N4245}
\end{figure*}

\addtocounter{figure}{-1}
\addtocounter{subfigure}{1}

\clearpage
\begin{figure*}
\begin{center}
  \includegraphics[angle=0, width=0.95\linewidth]{NGC4274_maps.ps2}
\end{center}
\caption{}
\label{fig:N4274}
\end{figure*}

\addtocounter{figure}{-1}
\addtocounter{subfigure}{1}

\clearpage
\begin{figure*}
\begin{center}
  \includegraphics[angle=0, width=0.95\linewidth]{NGC4293_maps.ps2}
\end{center}
\caption{}
\label{fig:N4293}
\end{figure*}

\addtocounter{figure}{-1}
\addtocounter{subfigure}{1}

\clearpage
\begin{figure*}
\begin{center}
  \includegraphics[angle=0, width=0.95\linewidth]{NGC4314_maps.ps2}
\end{center}
\caption{}
\label{fig:N4314}
\end{figure*}

\addtocounter{figure}{-1}
\addtocounter{subfigure}{1}

\clearpage
\begin{figure*}
\begin{center}
  \includegraphics[angle=0, width=0.95\linewidth]{NGC4369_maps.ps2}
\end{center}
\caption{}
\label{fig:N4369}
\end{figure*}

\addtocounter{figure}{-1}
\addtocounter{subfigure}{1}

\clearpage
\begin{figure*}
\begin{center}
  \includegraphics[angle=0, width=0.95\linewidth]{NGC4383_maps.ps2}
\end{center}
\caption{}
\label{fig:N4383}
\end{figure*}

\addtocounter{figure}{-1}
\addtocounter{subfigure}{1}

\clearpage
\begin{figure*}
\begin{center}
  \includegraphics[angle=0, width=0.95\linewidth]{NGC4405_maps.ps2}
\end{center}
\caption{}
\label{fig:N4405}
\end{figure*}

\addtocounter{figure}{-1}
\addtocounter{subfigure}{1}

\clearpage
\begin{figure*}
\begin{center}
  \includegraphics[angle=0, width=0.95\linewidth]{NGC4425_maps.ps2}
\end{center}
\caption{}
\label{fig:N4425}
\end{figure*}

\addtocounter{figure}{-1}
\addtocounter{subfigure}{1}

\clearpage
\begin{figure*}
\begin{center}
  \includegraphics[angle=0, width=0.95\linewidth]{NGC4596_maps.ps2}
\end{center}
\caption{}
\label{fig:N4596}
\end{figure*}

\addtocounter{figure}{-1}
\addtocounter{subfigure}{1}

\clearpage
\begin{figure*}
\begin{center}
  \includegraphics[angle=0, width=0.95\linewidth]{NGC4698_maps.ps2}
\end{center}
\caption{}
\label{fig:N4698}
\end{figure*}

\addtocounter{figure}{-1}
\addtocounter{subfigure}{1}

\clearpage
\begin{figure*}
\begin{center}
  \includegraphics[angle=0, width=0.95\linewidth]{NGC4772_maps.ps2}
\end{center}
\caption{}
\label{fig:N4772}
\end{figure*}

\addtocounter{figure}{-1}
\addtocounter{subfigure}{1}

\clearpage
\begin{figure*}
\begin{center}
  \includegraphics[angle=0, width=0.95\linewidth]{NGC5448_maps.ps2}
\end{center}
\caption{}
\label{fig:N5448}
\end{figure*}

\addtocounter{figure}{-1}
\addtocounter{subfigure}{1}

\clearpage
\begin{figure*}
\begin{center}
  \includegraphics[angle=0, width=0.95\linewidth]{NGC5475_maps.ps2}
\end{center}
\caption{}
\label{fig:N5475}
\end{figure*}

\addtocounter{figure}{-1}
\addtocounter{subfigure}{1}

\clearpage
\begin{figure*}
\begin{center}
  \includegraphics[angle=0, width=0.95\linewidth]{NGC5636_maps.ps2}
\end{center}
\caption{}
\label{fig:N5636}
\end{figure*}

\addtocounter{figure}{-1}
\addtocounter{subfigure}{1}

\clearpage
\begin{figure*}
\begin{center}
  \includegraphics[angle=0, width=0.95\linewidth]{NGC5689_maps.ps2}
\end{center}
\caption{}
\label{fig:N5689}
\end{figure*}

\addtocounter{figure}{-1}
\addtocounter{subfigure}{1}

\clearpage
\begin{figure*}
\begin{center}
  \includegraphics[angle=0, width=0.95\linewidth]{NGC5953_maps.ps2}
\end{center}
\caption{}
\label{fig:N5953}
\end{figure*}

\addtocounter{figure}{-1}
\addtocounter{subfigure}{1}

\clearpage
\begin{figure*}
\begin{center}
  \includegraphics[angle=0, width=0.95\linewidth]{NGC6501_maps.ps2}
\end{center}
\caption{}
\label{fig:N6501}
\end{figure*}

\addtocounter{figure}{-1}
\addtocounter{subfigure}{1}

\clearpage
\begin{figure*}
\begin{center}
  \includegraphics[angle=0, width=0.95\linewidth]{NGC7742_maps.ps2}
\end{center}
\caption{}
\label{fig:N7742}
\end{figure*}

\clearpage
\appendix
\section{Description of individual galaxies}
\label{sec:galaxies_notes}
Here, we briefly comment on the structures observed and provide some relevant
references in the literature for each galaxy.

\begin{description}

\item[\bf NGC 1056] displays a complex stellar velocity field, due to the
presence of significant amounts of dust. The inner parts, however, might reveal
the presence of an inner disk. Unlike other cases in the \sauron\ survey, this
feature is not accompanied by an anti-correlation with the $h_3$ parameter. The
velocity dispersion is also patchy, although a central dip can be recognised.
These properties contrast with a smooth and regular ionised-gas distribution and
kinematics. The \oiii/\hbeta\ ratio is low ($\le$1) all across the main disc of
the galaxy. This is in agreement with previous work in the literature that had
classified NGC\,1056 as a \hii\ galaxy \citep{veilleux95}.\looseness-2

\item[\bf NGC 2273] is a well-studied double barred galaxy \citep{fwm00}. As
expected in a low-inclined barred system, the \sauron\ stellar velocity field
is misaligned with respect to the bar major axis. The zero-velocity curve
displays a twist in the central $\approx$3\arcsec. This velocity feature is also
accompanied by a central dip in the velocity dispersion maps and anti-correlates
with the central $h_3$ values, suggesting the presence of a nuclear stellar
disc. The \oiii\ and \hbeta\ ionised-gas maps show very similar distributions
and kinematics. The \oiii/\hbeta\ ratio is enhanced in the centre of the galaxy,
in agreement with the Seyfert 2 classification of its nuclear emission
\citep{ho97}. Towards the end of the main bar, however, the ratio is lower,
which is consistent with the presence of \hii\ regions \citep{gptvr97}.

\item[\bf NGC 2844] is the brightest member of a group of galaxies including
also NGC\,2852 and NGC\,2853 \citep{gmcp00}. The stellar kinematics display 
regular velocity and velocity dispersion fields. The \hbeta\ and \oiii\ flux
maps reveal a ring-like structure, consistent with a dust ring seen in the
unsharp-masking.  The \oiii/\hbeta\ ratio and the gas velocity dispersion are
low along the ring, which it is indicative of ongoing star formation. The galaxy
is classified as an \hii\ galaxy in NED. The \hbeta\ and \oiii\ velocity fields
are both regular and similar to the stellar velocity map.

\item[\bf NGC 3623] is a barred Sa galaxy and a member of the Leo Triplet. While
there are strong indications that the two other members of the triplet
(NGC\,~3627 and NGC\,3628) are interacting, NGC\,3623 appears undisturbed
\citep{cemd98}. This galaxy contains a prominent dust lane at
$\approx$15\arcsec\ East of the nucleus. The stellar velocity map is twisted
with respect to the reconstructed intensity image. The $h_3$ parameter
correlates, in the outer parts, with the stellar velocity field,
giving kinematic confirmation of the presence of a large-scale bar \citep{ba05}.
A central disc is also present in the inner parts of the galaxy (see also
\citealt{as05}). Kinematic signatures of this disc can be seen in the stellar
velocity map, in the drop of velocity dispersion, and in the anti-correlation
of the $h_3$ parameter in the inner parts. The ionised gas shows a patchy
morphology with a highly twisted zero-velocity curve and some indication of
the presence of spiral arms. An \halpha\ image of the galaxy is shown in
\citet{hameed05}.\looseness-2

\item[\bf NGC 4220] is an almost edge-on galaxy ($i\approx85\degr$) that forms a
non-interacting pair with NGC\,4218 at $15\arcmin$ \citep{gmcp00}. Our
unsharp-masked image displays a prominent dust lane at a distance of 5\arcsec\
north-east of the centre, which is probably part of a more extended dust ring.
The presence of the dust lane is also evident in the \oiii/\hbeta\ map, where
the ratio is very low at that location. The \sauron\ reconstructed intensity
image reveals a clear boxy bulge component \citep{ldp00}. The stellar velocity
field exhibits cylindrical rotation. Both the \oiii\ and \hbeta\ morphology are
very similar in the central parts, but whereas the \oiii\ flux steadily
decreases further out from the center, the \hbeta\ emission remains strong also
outside the centre, in particular towards East. The ionised-gas velocity fields
are consistent with that of the stars.

\item[\bf NGC 4235] is a nearly edge-on Seyfert 1 galaxy \citep{jdtt00}
in a non-interacting pair with NGC\,4246 at 12\arcmin. The stellar velocity
field displays regular rotation and evidence for the presence of a smaller
stellar disk in the inner parts. This is supported by a decrease in velocity
dispersion and $h_3$ anti-correlation in the centre of the galaxy. The \oiii\
emission is much more extended along the main disc of the galaxy than the
\hbeta\ emission, which is mainly confined near the centre of the galaxy.
The \oiii/\hbeta\ ratio is high in the centre, consistent with its
classification as a Seyfert galaxy. The ionised-gas velocity maps of both
\hbeta\ and \oiii, although patchy, are consistent with the stellar velocity
field. Near-infrared surface photometry of this galaxy is presented in
\citet{peletier99}. Long-slit kinematics can be found in \citet{corsini03}.

\item[\bf NGC 4245] is a barred galaxy with a prominent dust ring in the centre
\citep{es03}. The stellar velocity field is misaligned with respect to the bar
major axis. There is also evidence, in the stellar kinematics (V, $\sigma$,
$h_3$), for a central disc in the inner 5\arcsec. A circumnuclear ring is
particularly evident in the \hbeta\ flux map. The correspondingly low values of
the \hbeta\ velocity dispersion and \oiii/\hbeta\ ratio suggests that star
formation is taking place at the same location. 

\item[\bf NGC 4274] is a double barred galaxy \citep{sapg95,erwin04} with
significant amounts of dust in the inner parts. Despite the dust, the stellar
kinematics maps show regular rotation on the main disc component of the galaxy,
and also the presence of an inner fast rotating ring (detected in the V and
$\sigma$ maps). As in NGC\,4245, the signatures of the star-forming ring are
easily identifiable in the ionised-gas maps. An \halpha\ image is presented in
\citet{hameed05}.
 
\item[\bf NGC 4293] is a relatively highly-inclined galaxy with a large-scale
bar \citep{ls02}. The unsharp-masked image displays a strong dust lane passing
through the nucleus and another dust lane about 7\arcsec\ south of the centre.
No correspondence between the dust lanes and features in the ionised-gas maps is
found. We find no signs of ongoing star formation in the \oiii/\hbeta\ map, in
agreement with \citet{kk04}. The stellar kinematics is regular with no signs of
inner kinematic components. 

\item[\bf NGC 4314] is a low-inclined, well studied barred galaxy in the Virgo
cluster \citep{bsk96, pkpldn00}, mainly known for a nuclear star-forming ring in
the inner 10\arcsec. The stellar kinematics displays regular rotation around the
minor axis of the galaxy. The kinematics of the ionised gas is consistent with
that of the stars. The presence of the nuclear ring is clearly visible in the
\hbeta\ flux and velocity dispersion maps. The \oiii\ distribution, however, is
more concentrated in the centre of the galaxy. The different morphology of the
two emission lines enhances the signature of the star-forming ring in the
\oiii/\hbeta\ map.

\item[\bf NGC 4369] is a low-inclined galaxy that shows a small rotational
velocity amplitude in our stellar velocity field. The stellar kinematics is
significantly affected by dust, as seen from the patchiness of the velocity
dispersion map. The ionised-gas flux maps show evidence for two star-forming
regions at both sides of the galactic centre \citep{ust98}. The \oiii/\hbeta\
map displays overall very low values overall. Near-infrared surface photometry
of this galaxy is presented in \citet{peletier99}, and \halpha\ imaging can be
found in \citet{hameed05}.

\item[\bf NGC 4383] is a starburst galaxy in the Virgo cluster \citep{rwk99}.
The stellar kinematics show a complex picture with no clear sense of rotation in
the inner parts, possibly due to the presence of significant amounts of dust. The
\sauron\ ionised-gas maps are also difficult to interpret, although they are
consistent with \halpha\ observations by \citet{kk04}, which reveal a bi-conical
filamentary structure suggestive of a starburst outflow. The galaxy also
displays enhanced star formation \citep{kk04}.\looseness-2

\item[\bf NGC 4405] is a Sa galaxy with no prominent spiral structure, but with
evidence of central dust lanes. The stellar velocity map is regular. Although
there is a drop in velocity dispersion at the centre of the galaxy, no evidence
for an inner component is found in either the stellar velocity or $h_3$ maps.
The \hbeta\ gas map displays several {\it clumps} around the centre, as opposed
to the \oiii\ map where the distribution is smoother. Both the \hbeta\ and
\oiii\ velocity maps show regular rotation, consistent with that of the stars.
Ongoing star formation is confined in the central dust disc seen in the
unsharp-masked image \citep{kk04}.

\item[\bf NGC 4425] is a highly inclined galaxy without much dust. The \sauron\
reconstructed intensity image reveals a boxy bulge component \citep{ldp00}. The
stellar velocity field exhibits cylindrical rotation. The stellar velocity
dispersion is generally low ($\le$100 \kms). The ionised-gas maps show only
patchy traces of emission.\looseness-2

\item[\bf NGC 4596] is a non-interacting strongly barred galaxy \citep{gkm99}.
The \sauron\ stellar velocity field displays regular rotation along an axis
misaligned with respect to the photometric major axis. There is evidence for a
disc-like component in the inner 5\arcsec\ region, that it is also apparent in
the velocity dispersion and $h_3$ maps. The distribution and kinematics of both
\hbeta\ and \oiii\ are aligned with the stellar kinematic major axis.
\citet{gkm99} , using the Tremaine-Weinberg method, reported that NGC\,4596 has
a fast bar.

\item[\bf NGC 4698] is a low-luminosity Seyfert 2 galaxy \citep{ho95} that is
thought to have experienced a major or intermediate-mass merger
(\citealt{bertola99}, but see also \citealt{sarzi00}). The \sauron\ stellar
velocity field displays regular rotation around the minor axis. The inner parts
however (i.e. 5\arcsec) appear to be rotating perpendicularly to the main body
of the galaxy, consistent with a nuclear disc in our unsharp-masked image
\citep[see also][]{pizzella02}. We note, however, that no clear evidence for
such decoupled inner component is found in either the velocity dispersion or
$h_3$ maps. This result is consistent with long-slit results from
\citet{bertola99}. The ionised-gas distribution is rather regular, with \hbeta\
being more extended than \oiii. The velocity fields of the ionised-gas are
consistent with that of the stars. \citet{kk04}, from \halpha\ observations,
reported a low of level star formation across the disc.

\item[\bf NGC 4772] is a galaxy located at the boundaries of the Virgo cluster
\citep{hjbbm00}. A strong dust lane is seen to the East. The stellar kinematics
display regular rotation along the major axis of the galaxy. The unsharp-masked
image reveals a small dusty disc in the centre of the galaxy, the orientation of
which seems consistent with that of the ionised gas in the central 5\arcsec. In
these regions both the \hbeta\ and \oiii\ lines suggest ordered rotation for the
gas, almost in the opposite sense of the stars, in agreement with \citet{hjbbm00}.
The ionised-gas emission also extends beyond the central regions, in particular
for the \oiii\ emission. The observed morphology is consistent with the
narrow-band observations of \citet{hjbbm00}, suggesting that the gas emission
observed at the north-eastern and south-western corner of the \sauron\ field
comes from an outer ring of gas on the main galactic-disk plane. The \oiii\
kinematics is more complex at larger radii, but suggests a warping of the
direction of maximum rotation towards the galaxy major axis.\looseness-2

\item[\bf NGC 5448] is an active barred galaxy \citep{ho97,eskridge02}. The
stellar kinematics display a regular disc, and signatures of an inner fast
rotating component in the central $\approx$5\arcsec. The \hbeta\ distribution
is more concentrated than the \oiii. The ionised-gas kinematics features an
'S-shaped' zero-velocity curve, suggestive of gas radial motions. The
\oiii/\hbeta\ map shows a drop in the centre of the galaxy, with the same
extent and location of the inner component seen in the stellar kinematics and
dust disc in the unsharp-masked image. We refer the reader to \citet{fathi05}
for a detailed study of this galaxy using \sauron. 

\item[\bf NGC 5475] is an isolated, nearly edge-on galaxy \citep{bp94}. The
\sauron\ stellar kinematics exhibits a fast rotating stellar disc aligned with
the photometric major axis of the galaxy (see also \citealt{fbpv03}). The \oiii\
emission, however, extends further out and along the galaxy minor axis. Although
disturbed, the ionised-gas kinematics further suggests evidence for polar
rotation along the major axis of the galaxy.

\item[\bf NGC 5636] is a strongly barred galaxy in a non-interacting pair with
NGC\,5638 at 2\arcmin. This galaxy is the smallest and has the lowest apparent
magnitude of our sample (m$_B$ = 13.7, from RC3). The direction of the maximum
stellar rotation is misaligned with respect to the bar major axis. The stellar
velocity dispersion is one of the lowest in the sample. In a large fraction of
the \sauron\ map the values represent an upper limit. The ionised-gas is
concentrated around the centre of the galaxy, where very low values of the
\oiii/\hbeta\ ratio suggest ongoing star formation.
 
\item[\bf NGC 5689] is an almost edge-on barred galaxy with a box-shaped bulge
\citep{ldp00}. It is the brightest non-interacting member of a group formed by
NGC\,5682, NGC\,5683, and NGC\,5693 \citep{gmcp00}. The unsharp-masked image
displays a dust lane $\approx$5\arcsec\ south of the nucleus. The stellar
kinematics shows prominent signatures of an inner, fast-rotating disc component
(i.e. central increase in rotational velocity, velocity dispersion drop, and
$h_3-V$ anti-correlation). See also \citet{fbpv03}. The ionised-gas shows an
elongated disc-like distribution and fast rotation.\looseness-2

\item[\bf NGC 5953] is a Seyfert 2 galaxy closely interacting with NGC\,5954
\citep{res93,gp96}. The \sauron\ stellar kinematics displays large-scale
rotation and a kinematically decoupled component in the central 10\arcsec. The
\oiii\ ionised gas presents an elongated distribution towards NGC\,5954 (i.e. NE
of  NGC\,5953 centre), which is associated with radio emission
\citep{jenkins84}, whereas the \hbeta\ distribution is symmetric around the
galactic centre and it is consistent with a dust ring seen in the unsharp-masked
image. The ionised-gas kinematics of the central component follows that of the
stars in the inner parts, and it is in good agreement with that of
\citet{hernandez03} from Fabry-Perot observations of the \nii\ emission-line.
The \oiii/\hbeta\ ratio map exhibits very low values in the location of the
circumnuclear region and it is enhanced in the very centre of the galaxy. Our
results are similar to those measured by \citet{yyktm93} in the inner region
(i.e. \oiii/\hbeta=0.14-0.41), and are consistent with a star-forming ring
surrounding an active galactic nucleus. The combined observations presented
here suggests that we might be witnessing the early stages in the formation of
a kinematically decoupled component. Further analysis including stellar
populations is, however, necessary to confirm this hypothesis.

\item[\bf NGC 6501] is galaxy that is part of a group formed by NGC\,6467, 
NGC\,6495, PGC\,61102 and NGC\,6500 \citep{gmcp00}. The \sauron\ stellar
velocity field displays regular rotation along the major axis of the galaxy. The
stellar velocity dispersion map shows the highest value in our sample
($\approx$270 \kms) and a steep gradient decreasing towards the outer parts. No
ionised-gas was detected in this galaxy, in agreement with earlier work
\citep{cid04}.\looseness-2

\item[\bf NGC 7742] is a well-known face-on Seyfert galaxy hosting a prominent 
star-forming ring surrounding a bright nucleus (see Paper II). Despite the low
inclination, the stellar kinematics show overall rotation at large-scale, but
also in the inner parts. The ring structure is evident in both the \hbeta\ and
\oiii\ distribution and velocity dispersions. The \oiii/\hbeta\ map displays
very low values at the location of the ring, also visible in the unsharp-marked
image, suggestive of current star formation. The ionised-gas counter-rotates
with respect to the stars. 

\end{description}

\label{lastpage}
\end{document}